\documentclass[iop]{emulateapj}
\begin{document}

\title{Constraining the Physical Conditions in the Jets of $\gamma$-ray Flaring
Blazars using Centimeter-Band Polarimetry and Radiative Transfer Simulations.
II. Exploring Parameter Space and Implications}
\author{Philip A. Hughes, Margo F. Aller, and Hugh D. Aller}
\affil{Astronomy Department, University of Michigan, Ann Arbor, MI 48109-1107}
\email{phughes@umich.edu, mfa@umich.edu, haller@umich.edu}

\begin{abstract}
 We analyze the shock-in-jet models for the $\gamma$-ray flaring blazars
0420-014, OJ~287, and 1156+295 presented in \citet[][Paper~I]{all14},
quantifying how well the modeling constrains internal properties of the flow
(low energy spectral cutoff, partition between random and ordered magnetic
field), the flow dynamics (quiescent flow speed and orientation), and the
number and strength of the shocks responsible for radio-band flaring. We
conclude that well-sampled, multifrequency polarized flux light curves
are crucial for defining source properties. We argue for few, if any, low
energy particles in these flows, suggesting no entrainment and efficient
energization of jet material, and for approximate energy equipartition
between the random and ordered magnetic field components, suggesting
that ordered field is built by non-trivial dynamo action from the random
component, or that the latter arises from a jet instability that preserves
the larger-scale, ordered flow.  We present evidence that the difference
between orphan radio-band (no $\gamma$-ray counterpart) and non-orphan
flares is due to more complex shock interactions in the latter case.
\end{abstract}

\section{Introduction}\label{intro}
 For most of the history of broadband studies of active galactic nuclei
(AGN) it has been the paradigm that the higher the energy of the emission,
the closer to the central engine that emission arises, with the highest
energies coming from the immediate environment of the supermassive black
hole and accretion disk \citep{mar06}. While this is likely to be true
of the quiescent emission, it has become clear over the last few years
that {\it flare} emission at high energies is {\it not} situated close the
central engine, but is, at least approximately, cospatial with the region
within which VLBI imaging reveals stationary or propagating knots in the
radio waveband.  Evidence for an association between high energy
(specifically GeV) and radio flaring is summarized in \citet[][hereafter
Paper~I]{all14}, and rests primarily on the observation that $\gamma$-ray
flaring always occurs at or near to the rise portion of a radio-band event
\citep[first noted by][]{vt95,vt96}, and component ejections seen in 43~GHz
VLBI monitoring data are temporally associated at a statistically significant
level with $\gamma$-ray flares \citep[e.g.,][]{jor01}. The EGRET results
have been confirmed by recent studies using the wealth of data from Fermi.
For example, \citet{agu11} have presented strong evidence that $\gamma$-ray
flaring in OJ~287 is located $>14$ pc from the central engine, \cite{jor13}
present evidence that $\gamma$-ray and low-frequency events are cospatial
in the source 3C~454.3, while \citet{ack14} tentatively associate a knot
seen in 43~GHz imaging of PKS 1222+216 with a $\gamma$-ray flare, also
putting the event in the parsec-scale flow.  However, statistical studies
based on cross correlations of $\gamma$-ray and radio-band light curves
have been less clear-cut in providing evidence that the same disturbance
is responsible for activity in both bands: \citet{max14} find that
only one out of 41 sources (using 4 yr of 15 GHz observations from the
Owens Valley Radio Observatory 40-m monitoring program) show correlated
$\gamma$-ray-radio-band activity at a significance greater than $3\sigma$
(and only 4 sources have any statistically significant correlation); it is
likely that the correlation analysis is limited by uneven sampling, and the
almost continuous activity exhibited by most sources.  More promisingly,
\citet{kov09} find that radio jets are more active within months of
Fermi-detected $\gamma$-ray emission. While the statistical studies are
inconclusive, the association of $\gamma$-ray and radio-band events, on a
case-by-case basis, supports the view that the $\gamma$-ray mission arises
at parsec-scales, and motivation for the modeling discussed here.

 On the premise of a strong connection between the radio-band and
$\gamma$-ray flares, detailed modeling of the former provides a powerful
tool with which to help understand the origin of the high energy emission,
and thus the flow dynamics and particle kinematics of relativistic flows.
A wealth of data -- multifrequency light curves in total and polarized
flux from single dish observations, and component fluxes, ejection times,
and speeds from VLBI observations -- coupled with radiation transfer
calculations for shocks propagating in a diverging flow -- the widely
accepted picture for the interpretation of radio-band data -- can provide a
detailed picture of the flow structure, dynamics, state and orientation.
That provides a framework within which to understand the physical origin
of the $\gamma$-ray flares, and can potentially shed light on why some
radio flares have no high energy counterpart, and vice versa.

 Motivated by these considerations, we extended our original shock modeling
\citep{hug85,hug89a,hug89b} to include arbitrary shock obliquity with
respect to the flow direction \citep{hug11}, and used that study as the
basis of models for three sources which displayed cotemporal radio-band and
$\gamma$-ray flares, and which had sufficiently well-defined structure in
their total and polarized light curves that model-fitting to the several
subflares that comprise each distinct outburst was viable (Paper~I). The
selected sources were the QSO 0420-014, the BL~Lac object OJ~287, and the
FSRQ 1156+295. We found that the general structure and spectrum of both
the total and polarized flux light curves could be reproduced with forward
moving shocks, preferentially oriented transversely to the flow direction,
with OJ~287 being distinct in needing a significantly oblique shock structure,
and very low cutoff energy for the radiating particle distribution.

 The modeling was able to provide an estimate of the angle at which the
observer views the flow: that parameter is very well-constrained because
of the sensitivity of the level of polarized flux to the direction from
which the source is viewed. The source parameters, including the viewing
angle, are determined independently of VLBI observations, and yet provide
component speeds and viewing angle in agreement with those found from
VLBI data for the sources modeled, validating the modeling. A
particular value of the modeling is that it enables an estimation of these
(and other) source properties, for sources for which no VLBI data exist.

It was noted in Paper~I that one of the best-determined source parameters
is the observer's angle of view with respect to the jet axis, as this
plays a major role in establishing the level of polarized emission, and
small changes in the viewing angle have a large impact on the amplitude of
$P$. The scope of Paper~I did not allow a full discussion of the complete
set of model parameters, nor an exploration of the impact of changing
those parameters on the model light curves, and thus a quantitative
assessment of how well-constrained each parameter is. It is the purpose
of the current paper to discuss the choice of parameter values, quantify
how well each is determined, and illustrate the changes to the model light
curves that result from variation in those parameters to which the modeling
is most sensitive. Section~\ref{modparam} provides a broad overview of all
parameters needed to specify a source model, while Section~\ref{explore}
contains a detailed analysis of the roles played by the low-energy cutoff in
the radiating particle spectrum, the axial magnetic field strength, the flow
Lorentz factor, the observer's angle of view with respect to the jet axis,
and the azimuthal angle of view. The azimuthal angle is not significant
for transverse shocks, but it is potentially important if the shock is
highly oblique. Section~\ref{orphan} explores how changing the number
and strength of shocks contributing to a single radio outburst impacts
the light curve in the context of the `orphan flare' in the light curve
of 1156+295. It was noted in Paper~I that the modeling ignores retarded
time effects, since they produce negligible changes in the light curves;
justification for that is presented in an appendix.

\section{Model Parameters}\label{modparam}
 All computations have been performed in a volume of $61\times 61\times
600$ cells. The lateral extent of the volume is sufficient to allow
the modeling of a diverging flow, while the length can accommodate multiple
shocks within the flow at one time. The additional computational resources
needed to explore a larger volume would limit the range of sources,
events, and parameters explored, and is unnecessary judged by the absence
of artifacts associated with limited resolution in the light curves.

 Each simulation has used a jet with opening half-angle $2.4\arcdeg$, with
initial radius $1/12$th the lateral extent of the computational volume.
These were chosen to optimize the use of the volume: adiabatic expansion of a
component reduces the total and polarized flux to close to the jet quiescent
level by the time that component has traversed the volume. Models are
insensitive to the opening angle to the extent that an increase (decrease)
in the angle causes a more (less) rapid fall of flux as the component
propagates; this translates into a smaller (larger) number of computational
time steps being mapped into the observed event duration during modeling,
thus using the volume less effectively, but not changing the general trends
with time and frequency seen in the model light curves. Changing the flow
opening angle will change the distribution of flow velocity vectors seen by
an observer placed at some viewing angle from the flow axis, but for quite
well-collimated flows, the impact of this on the light curves will be subtle,
and matching models to data provides little constraint on the opening angle.

Computations are performed in dimensionless units, with particle density,
magnetic field strength, line-of-sight source extent and observational
frequency scaled with fiducial values. Associating a particular model event
with data then establishes a length (through association with structure
on a VLBI map) or a time (through association with a light curve flare
duration). With absolute time (length) established, the speed of light
then establishes an absolute length (time). However, that still leaves
particle and energy densities unknown, and thus a model is scaled to
match the peak model flux at the highest frequency to the peak flux of the
modeled flare. For the same reason, opacity (a function of line-of-sight
source extent, particle and field densities, and observational frequency)
is unknown, and is incorporated using the same scaled quantities, subject
to an overall arbitrary scaling. Effectively, a target optical depth
for a representative line-of-sight, and fiducial values of all pertinent
physical quantities, are set, and the local opacity is scaled throughout
the volume accordingly.  In modeling, this parameter is adjusted to give
the best match to the spectral behavior of the event at peak flux.

The radiating particle energy distribution is assumed to follow a power-law
in Lorentz factor, $n\left(\gamma\right)d\gamma=n_o\gamma^{-\delta}d\gamma,\
\gamma>\gamma_i$. As noted below, previous studies suggest a fairly flat
spectrum, and thus that there must be an upper cutoff to keep the total
energy in the particle distribution finite. However, that cutoff is assumed
to be well above the energies seen by GHz-band University of Michigan
Radio Observatory (UMRAO) data, and is not pertinent to the GHz-band
modeling. We plausibly, but arbitrarily, assume that the particles seen
through their synchrotron emission in the central UMRAO band (8~GHz) have
a fiducial Lorentz factor $\gamma_c=10^3$.  This implies a magnetic field
strength in the emitting region through $\gamma_c^2\sim \nu\left({\rm
8\,GHz}\right)/\nu_G$, where $\nu_G$ is the particle gyrofrequency. A
different choice of $\gamma_c$ would imply a different field strength,
and concomitant change in flux density. However, as we are concerned
only with the spectral and temporal form of the light curves, which are
subject to an arbitrary scaling, this has no impact on the modeling.
The purpose of setting $\gamma_c$ is to establish the extent to which a
certain choice of $\gamma_i$ extends the spectrum to include particles
that can contribute significantly to internal Faraday effects.  Such an
approach assumes that no Faraday effects arise from a distribution of (cold)
thermal electrons within the jet, which is consistent with the most recent
studies, such as the analysis of circular polarization in the source PKS
B2126-158 by \citet{osull13}, who also summarize arguments against there
being a significant thermal electron content in such flows.  Additionally,
\citet{hov12} present a discussion of MOJAVE observations showing that
they are consistent with external Faraday screens being responsible for
the polarization behavior exhibited by {\it most} sources.

The power-law index of the radiating particle distribution is determined by
fixing the optically-thin frequency spectral index as $\alpha=0.25$, where
$\alpha$ is defined by $S\left(\nu\right)d\nu=S_0\nu^{-\alpha}d\nu$. This
value is chosen by inspection of the data for many UMRAO-observed sources,
which typically display a rather flat spectrum even in the optically-thin
state. Since the work of \citet{bk79} it has been conventional wisdom that
the flat optically-thin spectra of blazars are a consequence of summing the
multiple segments that make up a quiescent flow, each with slightly different
peak frequency. However, the first detailed shock models \citep{hug89b}
suggested that a flat spectrum is an intrinsic feature of these flows,
a truncation of the jet being needed for a good fit of the model to the
data. This is not surprising: as discussed by \citet{mar06}, self-similarity
cannot extend down to the region of jet formation. An observable jet starts
some distance from the central engine, possibly where particle acceleration
and field amplification occur at a recollimation shock \citep{caw13}.

As discussed in \citet{hug11}, three components of magnetic field have
been considered, two (random, and ordered helical) widely believed to play
a role in determining the character of the emitted radiation, and a third
(ordered axial) to provide a well-defined electric vector position angle
(EVPA) in the quiescent state. Of the two potentially dominant components,
a random field is established in each jet as set out in \citet{hug11}, while
the helical component discussed in that paper has not been incorporated
in the current modeling, as the previous study showed that a significant
contribution (when the helical component contributes of order half, or
more, of the total magnetic energy density) from such a field predicted
evolution of the EVPA not in agreement with UMRAO data. Modeling of the
sources 0420-014, OJ~287, and 1156+295 as discussed in Paper~I led to
the unexpected result that the axial magnetic field, originally thought
to be unimportant during outbursts, and set to provide a well-defined
quiescent EVPA, can play a significant role in determining the properties
of the polarized emission while a source is in outburst. The magnitude of
that is thus an important parameter, and is characterized by $\bar{B_z}$,
in units such that the average random field has unit magnitude: $<B_{\rm
ran}^2>=1$. Thus a flow with {\it only} the axial field would have a magnetic
energy density a fraction $\bar{B_z^2}$ that of the purely random case.

The Lorentz factor of the bulk, unshocked flow ($\gamma_f$), is specified
independently of the shock strengths, and is typically in the range 5-10.
The compression of each shock is specified as a fraction $\kappa<1$, so
that the passage of a shock compresses unit length to length $\kappa$.
For a relativistic equation of state and given shock obliquity this
compression uniquely determines the upstream and downstream Lorentz factors
in the frame of the shock transition \citep{hug11}. The speed of the shock
transition can then be calculated given the bulk Lorentz factor, once a
choice between reverse or forward shocks has been made. A shock system is thus
completely specified by choice of bulk Lorentz factor, shock obliquity,
shock sense (forward or reverse), and the compressions of the individual
shocks. Additionally, each shock has a certain start time and has a length
(extent of the shocked region) expressed as a fraction of the quiescent
flow length.

The remaining fundamental parameters are the polar ($\theta$) and azimuthal
($\phi$) angle of the observer. The jet axis provides a natural reference
for the polar angle ($\theta=0$), while the zero-point of the azimuthal
angle is arbitrary. For a transverse shock (i.e., with the shock normal
parallel to the jet axis) the azimuthal orientation of the observer will
play no role in what is seen. However, as a general rule, `transverse
shocks' are modeled with a slight offset (typically $1\arcdeg$), in part
as it is computationally convenient to avoid infinities in the analytic
forms used to compute speeds and deflections, but also because running a
model through a range of azimuthal angles can then give insight into the
sensitivity of the model to slight variations in obliquity.

Formally, an additional parameter, the orientation of the shock normal with
respect to an arbitrary azimuthal reference direction, $\psi$, can also
be specified. However, that is degenerate with the observer's azimuthal
viewing angle, $\phi$, and need not be considered separately. As noted in
\citet{hug11} and Paper~I, the azimuthal angle does not play a major
role in determining the {\it general form} of the total and polarized flux
light curves, but, in particular for very oblique shocks, the appropriate
choice of angle can be important for reproducing certain details of the
data -- such as the relative percentage polarization of shocks within a
flare envelope.

In summary, the fundamental parameters for exploration are those
characterizing the internal state of the quiescent flow, the bulk dynamics
and orientation of that flow, and the attributes of the disturbances
(shocks) to that flow, giving rise to non-steady emission; namely
\begin{itemize}
\item the low-energy cutoff of the radiating particle distribution,
   $\gamma_i$, and the axial magnetic field characterized by $\bar{B_z}$;
\item the bulk Lorentz factor ($\gamma_f$) and viewing angle ($\theta$,
   $\phi$) of the flow;
\item the obliquity ($\eta$), sense ({\bf F} or {\bf R}), compression
   ($\kappa$), start time, and length of each shock.
\end{itemize}

\section{Parameter Exploration}\label{explore}
The obliquity ($\eta$), sense ({\bf F} or {\bf R}), compression ($\kappa$),
start time, and length of each shock are adjusted as part of the modeling
process. This parameter set, in part, defines a distinct source model. The
obliquity, sense, compression and length of each segment of shocked flow
will depend on the (unspecified) process that originally set up the shocks,
while the start times establish a causal relationship between these events
and both evolving VLBI structures and flaring in other emission bands. Values
of these parameters for the models of sources 0420-014, OJ~287, and 1156+295
are presented and discussed in Paper~I. They are typical of values found
in earlier shock models of blazar flares \citep{hug89b,hug91}, except that
the recent modeling favors forward moving, rather than reverse, shocks.

On the other hand, the low-energy cutoff of the radiating particle
distribution, $\gamma_i$, the axial magnetic field, $\bar{B_z}$, the bulk
Lorentz factor ($\gamma_f$), and the viewing angle ($\theta$, $\phi$)
of the flow, are fundamental parameters of the jet/shock system. There
are implications for jet magnetohydrodynamics and particle acceleration
processes in the case of the cutoff, field and Lorentz factor, and
implications for the appearance of flares -- and potentially the relative
prominence of flares in different parts of the radio spectrum -- in the case
of all these parameters. It is thus important to understand how well these
quantities are established by the modeling. In particular, exploring how
models change as these parameters are varied will establish how different
these fundamental parameters are in sources with behavior broadly similar
to that of the modeled sources, but with differences in detail -- for
example, in the total flux density spectrum, or in the amplitude and
complex variations of the polarized flux.

We consider each of these parameters below, using the models of 0420-014
and 1156+295 (with transverse shocks) and OJ~287 (with oblique shocks) as
discussed in Paper~I. A `library' of light curves ($S$, $P$ and EVPA at
each of three frequencies) could be constructed for a single flare over
a grid of these parameters, and a sequence of shock obliquities, but a
well-sampled exploration of a 6-parameter space would lead to an unmanageable
set of complex plots, which would in fact provide little physical insight,
because a) most interest lies in the complex interaction of multiple events
-- indeed, the UMRAO data show that single shock models are physically
unrealistic; and b) the variation of a given parameter is often significant
in a particular way: for example, variation of the contribution of an
on-axis magnetic field significantly influences the degree and structure
exhibited by the percentage polarization, leaving the total flux density
largely unchanged. It is therefore more useful to focus on the consequences
of varying one parameter at a time, in the context of a representative model.

The resulting figures are complex (multiple panels, each containing multiple
curves -- one for each of the harmonically related frequencies modeled).
Overlaying a number of such displays leads to graphs of such density and
complexity as to be unreadable, while a mosaic of plots renders each too
small for the detail to be seen, and does nothing to facilitate appreciating
the changes that occur as a parameter is varied. To show the range of
behavior we display two sets of model curves only -- for each extreme of
the parameter range -- to highlight the consequence of parameter variation,
and animations are available in the online version of the journal that
enable to reader to step through, pause, and replay sequences of models.

\subsection{Low Energy Cutoff}\label{lowenergycutoff}
 In the absence of a `cold' plasma component, Faraday effects are associated
with the relativistic particles in the low energy part of the radiating
particle spectrum. Following the notation of \citet{jod77}, whose analysis
has formed the basis of our modeling since the work of \citet{hug89b},
such effects depend on the normalized rotativity
\begin{equation}
\zeta_v^* = \zeta_{v\alpha}^* \frac{\ln \gamma_i}{\gamma_i}
   \left(\frac{\gamma_c}{\gamma_i}\right)^{2\alpha+1} 
   \left(\frac{\nu}{\nu_c}\right)^{\alpha+1/2}
   \cot \vartheta,
\end{equation}
where $\zeta_{v\alpha}^*$ is a constant of order unity, and $\nu_c$ is the
emission frequency for particles of energy $\gamma_c$. For observations
at frequency $\nu\sim\nu_c$, a random field element oriented with respect
to the observer such that $\cot \vartheta\sim 1$, and, as noted above
$\alpha=0.25$, for the rotativity to play a significant role we need
\begin{equation}
\frac{\ln \gamma_i}{\gamma_i} \left(\frac{\gamma_c}{\gamma_i}\right)^{1.5} \gtrsim 1,
\end{equation}
or $\gamma_i \lesssim \gamma_c^{0.6}$. Given $\gamma_c=10^3$, this implies we
can expect to see significant Faraday effects for $\gamma_i$ less than
about $60$.

\begin{figure}
\includegraphics[scale=0.45,clip=true]{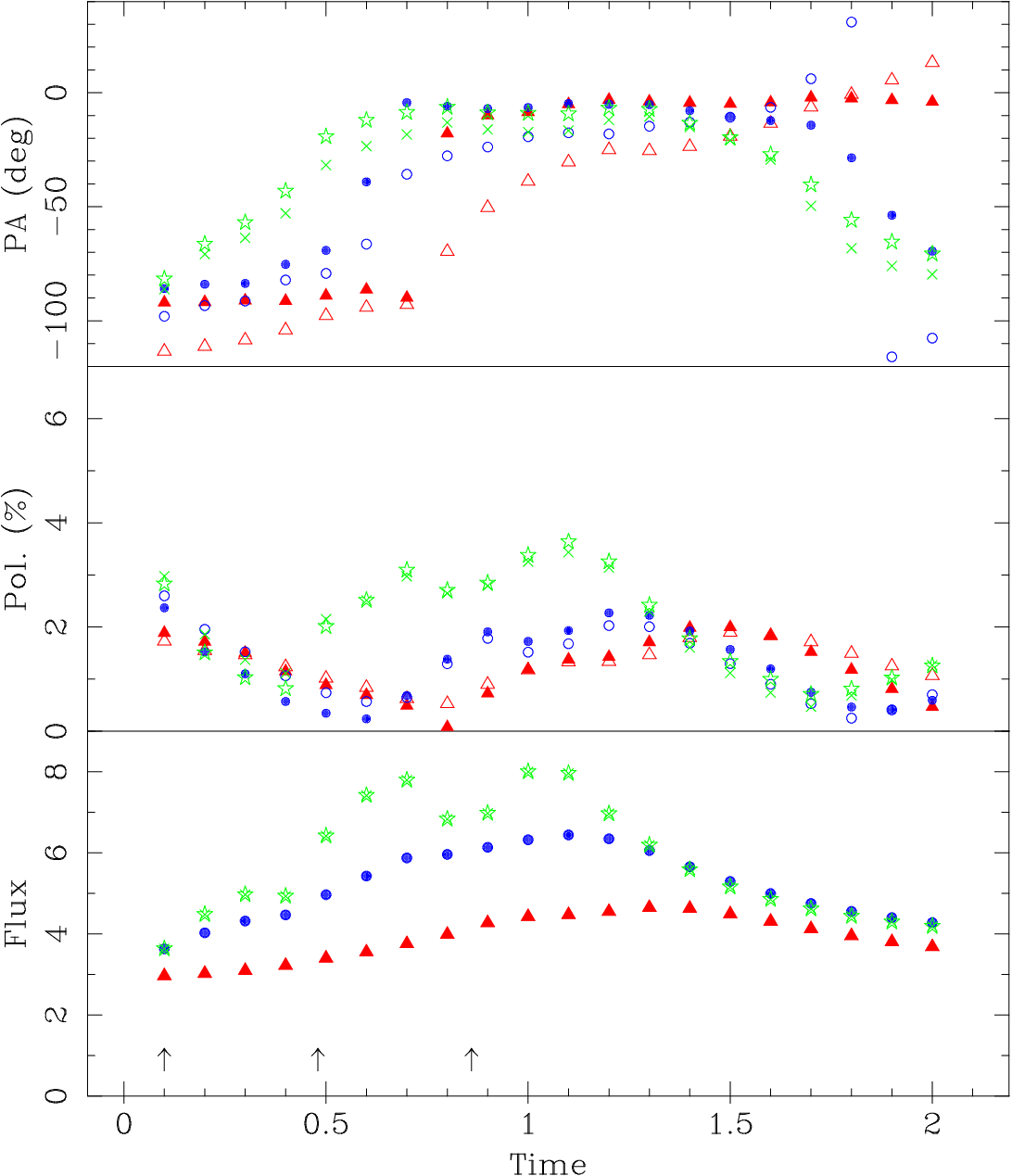}
\caption{The model for 0420-014 at two extremes of the low-energy spectral
cutoff: $\gamma_i=25$ is shown using unfilled triangles, circles and
crosses for the frequencies corresponding to the UMRAO frequencies
of 4.8, 8.0, and 14.5~GHz (the same convention adopted in Paper~I),
while $\gamma_i=75$ uses filled triangles, circles, and solid crosses.
Note that the flux curves do not change, and only one set of points is
evident in the bottom panel. The onset time of each shock is shown by
arrows in the bottom panel. An mpeg animation showing
in detail the variation of the light curves with change in model parameter
is available online.
}
\label{fig1}
\end{figure}

The parameter sequence shown in Figure~\ref{fig1} (animation online) spans
the range $25\le \gamma_i \le 75$, and is based on the model for 0420-014,
for which a value of $\gamma_i=50$ was adopted. 
\footnote{
Animations of model sequences are available at:
http://dept.astro.lsa.umich.edu/\~{ }phughes/GALLERY/APJ14/.}
The three panels are, from
bottom to top, the total flux density, the percentage polarization, and
the EVPA (position angle of the electric vector of the polarized emission)
at three harmonically related frequencies. In this and the following cases
(with one exception noted below) the figures show the models from the first
and last frames of the animation sequences, with the model for the lowest
value of the varied parameter depicted using unfilled triangles, circles
and crosses for the frequencies corresponding to the UMRAO frequencies of
4.8, 8.0, and 14.5~GHz (the same convention adopted in Paper~I), while the
model for the highest value of the parameter is shown with filled triangles,
circles, and solid crosses. Arrows in the bottom panel show the onset time
of each shock.

Changes in the percentage polarization are small; indeed, they are
perceptible only for $\gamma_i<50$. Below this value, lower values of
$\gamma_i$ lead to a slightly raised value of the percentage polarization
(tenths of a percent) at the lowest frequency (4.8~GHz) between the peaks
of the second and third shocks. The increased Faraday depth (significant
only at the lowest frequency) causes a slight rotation of the EVPA of
the emission from the most optically thick part of the source, changing
the percentage polarization integrated over the entire jet. However, this
effect is too small to play any role when confronting the data with models
designed only to reproduce general trends.

For the highest value of $\gamma_i$ explored here, there is a small
separation in the EVPAs with frequency at the initial time, which exhibits
no significant change between values of $65$ and $75$, indicating that
the small residual difference is opacity-related.  The separation of the
EVPA curves with frequency increases during the onset of the outburst
because of the opacity-induced delay in seeing the orthogonally-polarized
shock emission.  Quantitatively, this trend is not significantly influenced
by the value of $\gamma_i$.  Towards the end of activity, emission at the
highest frequency has become sufficiently optically-thin that the quiescent
flow begins to dominate again, and there is reversion towards the original
EVPA. Only very low cutoff values, and opacity higher than that adopted
in this model, can be expected to reveal the effects of the cutoff in the
flare light curves, as for example in the case of OJ~287 (Paper~I).

The sense of Faraday rotation is dependent on the sense of the magnetic
field component along the line-of-sight, and while a finite number of
randomly-oriented magnetic cells will lead to a dispersion in net rotation
over a set of realizations of that field, the effects we have discussed
above are associated with the ordered, axial field of the flow. In that sense,
the absence of an ordered field is degenerate with a higher value of the
cutoff energy.

\begin{figure}
\includegraphics[scale=0.45,clip=true]{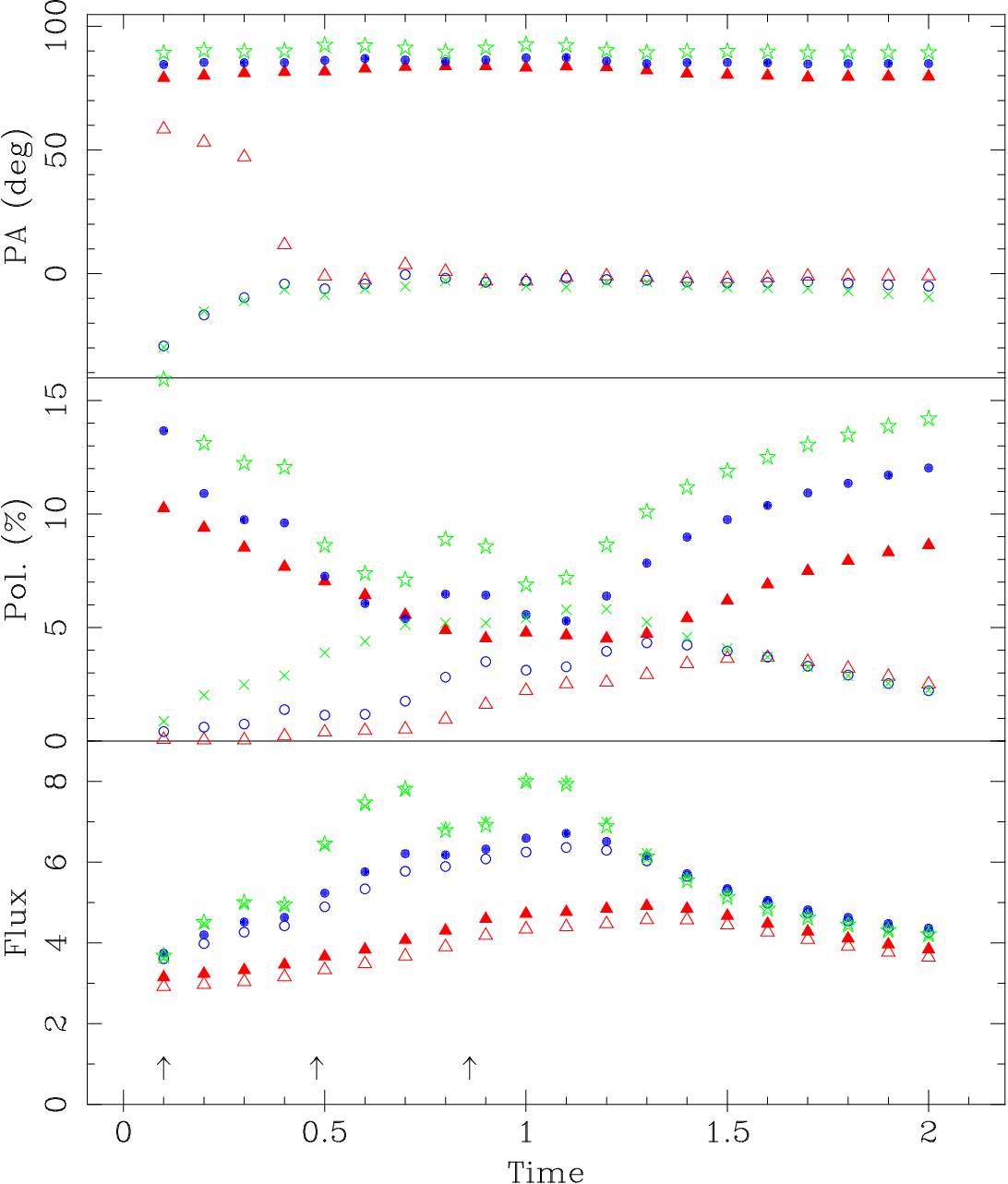}
\caption{The model for 0420-014 at two extremes of the mean field
contribution: $\bar{B_z}=0.05$ and $1.0$.  The symbols are as for
Figure~\ref{fig1}.  An mpeg animation showing in detail
the variation of the light curves with change in model parameter is
available online.
}
\label{fig2}
\end{figure}

\subsection{Axial Magnetic Field}\label{axifield}
We now discuss how sensitive the model fitting is to an ordered component
of the magnetic field.  The parameter sequence shown in Figure~\ref{fig2}
(animation online) spans the range $0.05\le \bar{B_z} \le 1.0$, and is
based on the model for 0420-014, for which a value of $\bar{B_z}=0.4$ was
adopted. (Note that in the animation the sample values are not uniformly
distributed in the adopted range.) The `fiducial' value corresponds to
a magnetic energy density $\bar{B_z^2}\sim 0.16$, so that the mean field
contributes less than 20\% of the energy density contributed by the random
component. Inspection of the parameter sequence reveals that a change in the
character of the polarized emission -- the percentage polarization and EVPA,
and their frequency and time dependence -- occurs at $\bar{B_z}\sim 0.5$,
at which point the mean field is contributing $\sim 20\%$ of the total
magnetic energy density.

The quite rapid switch to a high degree of polarization, and EVPA orthogonal
to that displayed during the shock-dominated, weak mean field case, for
values $\bar{B_z}\gtrsim 0.5$ is not surprising. As noted by \citet{hug11},
even small spurious Fourier components associated with generating a random
magnetic field by selecting random phases and amplitudes for the Fourier
transform of the magnetic vector potential can lead to significant spurious
polarized flux. However, this does lead to the question of how plausible
it is to invoke an ordered magnetic field to explain the amplitude and
detailed structure of blazar polarized flux light curves, if a modest
increase in the strength of that component can lead to features (very high
percentage polarization, for example) never seen in those same light curves.

In fact, the fundamental premise of our modeling -- that these flows are
turbulent -- suggests that the ordered field is not a direct manifestation
of the central black hole/accretion disk system to which the jet is tied,
but is generated {\it in situ} through a dynamo process \citep{eh91}.  Mean
field dynamos in sheared, turbulent flows are known to occur \citep{rk03},
and while the process has not been explored for a cylindrical geometry
(for which it can be speculated that the mean field takes the form of a
force-free flux tube), shearing box simulations \citep{y08} show the dominant
mean field to be in the sense of the flow -- which is along the axis in
the case of a jet.  It would then be expected that a mean field would grow,
and then saturate at an energy density some fraction of order unity that of
the kinetic energy density of the underlying turbulence; the latter would be
expected to be comparable to the energy density in the random field. This
is impossible to quantify, as there are inconsistencies between state
of the art analytic theory and simulations in this field, and it remains
impossible to apply the latter to realistic values of scale separation,
Reynolds number, etc. \cite[e.g.,][]{park13}. However, simulations such as
presented in that study strongly suggest such a saturation, as do recent
studies of relativistic turbulence \citep{zrake14}. In the context of
this interpretation, the importance of a weak ordered field in explaining
the light curves, while never dominating the light curves, is evidence
supportive of the scenario in which any large scale field grows through
dynamo action in a fundamentally turbulent flow.

\begin{figure}
\includegraphics[scale=0.45,clip=true]{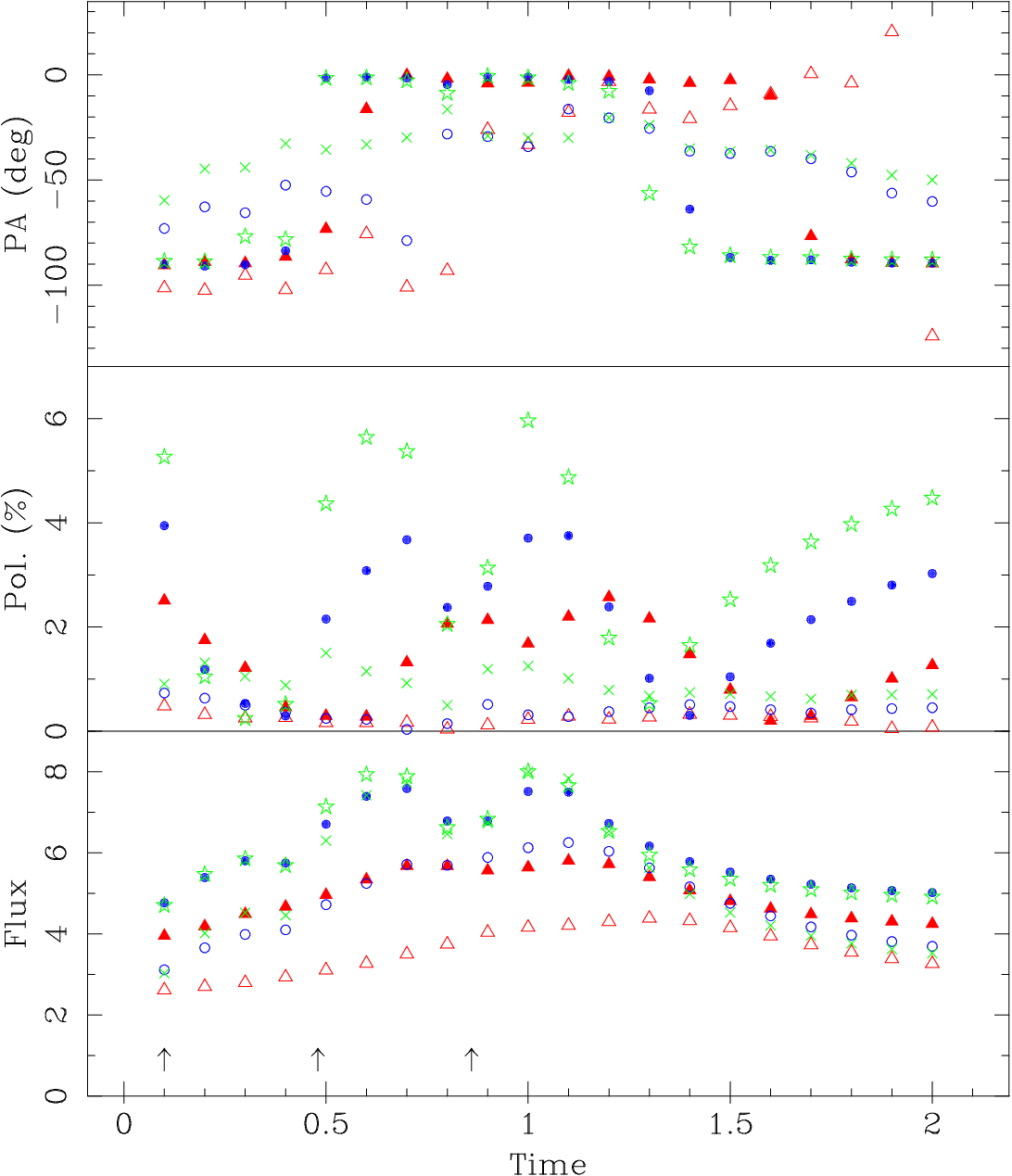}
\caption{The model for 0420-014 at two extremes of the flow Lorentz factor:
$\gamma_f=2.0$ and $43.0$. The symbols are as for Figure~\ref{fig1}.
An mpeg animation showing in detail the variation of the
light curves with change in model parameter is available online.
}
\label{fig3}
\end{figure}

\subsection{Bulk Lorentz Factor}\label{lorentzfactor}
The parameter sequence shown in Figure~\ref{fig3} (animation online)
spans the range $2.0\le \gamma_f \le 43.0$, and is based on the model for
0420-014, for which a value of $\gamma_f=5.0$ was adopted, and for which the
viewing angle is $4\arcdeg$. Each set of plots in the sequence increases
the Lorentz factor by $1.05$ relative to the speed of the previous case.
For fixed angle of view and observing frequency, a change in the Lorentz
factor leads to a change in the Doppler factor, and thus to a change in the
emission frame frequency.  The adopted value of the opacity was therefore
adjusted to compensate for this, so that the asymptotic spectral shape
(the spectral slope at the end of the flare, when close to quiescent)
was the same in each case. The opacity drops by a factor $\sim 2$ per
step in Lorentz factor through the first half of the sequence, and then
asymptotes to a value $\sim 0.024$ times the initial value, between Lorentz
factor values of $12.2$ and $16.7$; beyond this point in the sequence,
the viewing angle lies outside the critical ($1/\gamma$) cone of the flow.

The spectral shape during flaring changes along this sequence because the
Lorentz factor of the shocked flow differs from that of the quiescent flow,
the spectral characteristics of which are used to normalize the opacity. (The
quiescent flux rises along the sequence because the peak flux is normalized
to that of the modeled data, and the spectrum becomes shallower.) For
the model of 0420-014 the Lorentz factor of the shocked flow is $\sim 8$,
required to yield the necessary shock compression, in the forward shock
case. In the sequence under discussion, the Lorentz factor is varied,
but the shock compressions adopted for the 0420-014 model are retained
from case to case.  For flows slower (faster) than the fiducial case, a
given compression is achieved with a shocked flow with Lorentz factor $<8$
($>8$).  The adjustment to the opacity thus under-compensates during the
flare for the slower flows. As for the quiescent spectral shape, that of
the flare asymptotes at high flow speed.

For the smallest Lorentz factor of the sequence the percentage polarization
is very low -- rising just above 1\% at the highest frequency, and negligible
at the lower frequencies. It rises rapidly along the sequence, attaining
values in the range 2-4\% for the Lorentz factor of the 0420-014 model,
peaks with maximum value $\sim 6\%$ at the highest frequency for $\gamma_f
\sim 12$ and then declines slowly. As noted above, this corresponds to the
critical cone of the flow passing through the observer's line-of-sight;
at that angle the observer sees radiation emitted orthogonal to the flow
direction, which for a compression transverse to the flow will maximize
the polarization. That orientation also maximizes the projection of the
mean axial field on the observer's plane-of-sky. Below $\gamma_f \sim 12$
the rapid change in degree of polarization with Lorentz factor provides a
strong discriminant between models for data such as those for 0420-014,
while higher values of the Lorentz factor are excluded in this case by
both the high model percentage polarization in the flare-state, and in
quiescence, where the polarized emission is significantly influenced by
the axial mean field (seen in the emission frame at an angle of $37\arcdeg$
for $\gamma_f=43$).

For the lowest Lorentz factors the degree of polarization is low, the EVPA
is subject to substantial spread with frequency, and trends are difficult
to define.  By $\gamma_f\sim 3.5$ the polarization has risen to a level
that allows a well-defined EVPA, the trends in which then persist across
the sequence to the highest Lorentz factor explored.  The 4.8~GHz data
are sparse, but the 8.0 and 14.5~GHz data points and the model exhibit a
swing in EVPA by $\sim90\arcdeg$ that are in agreement.

\begin{figure}
\includegraphics[scale=0.45,clip=true]{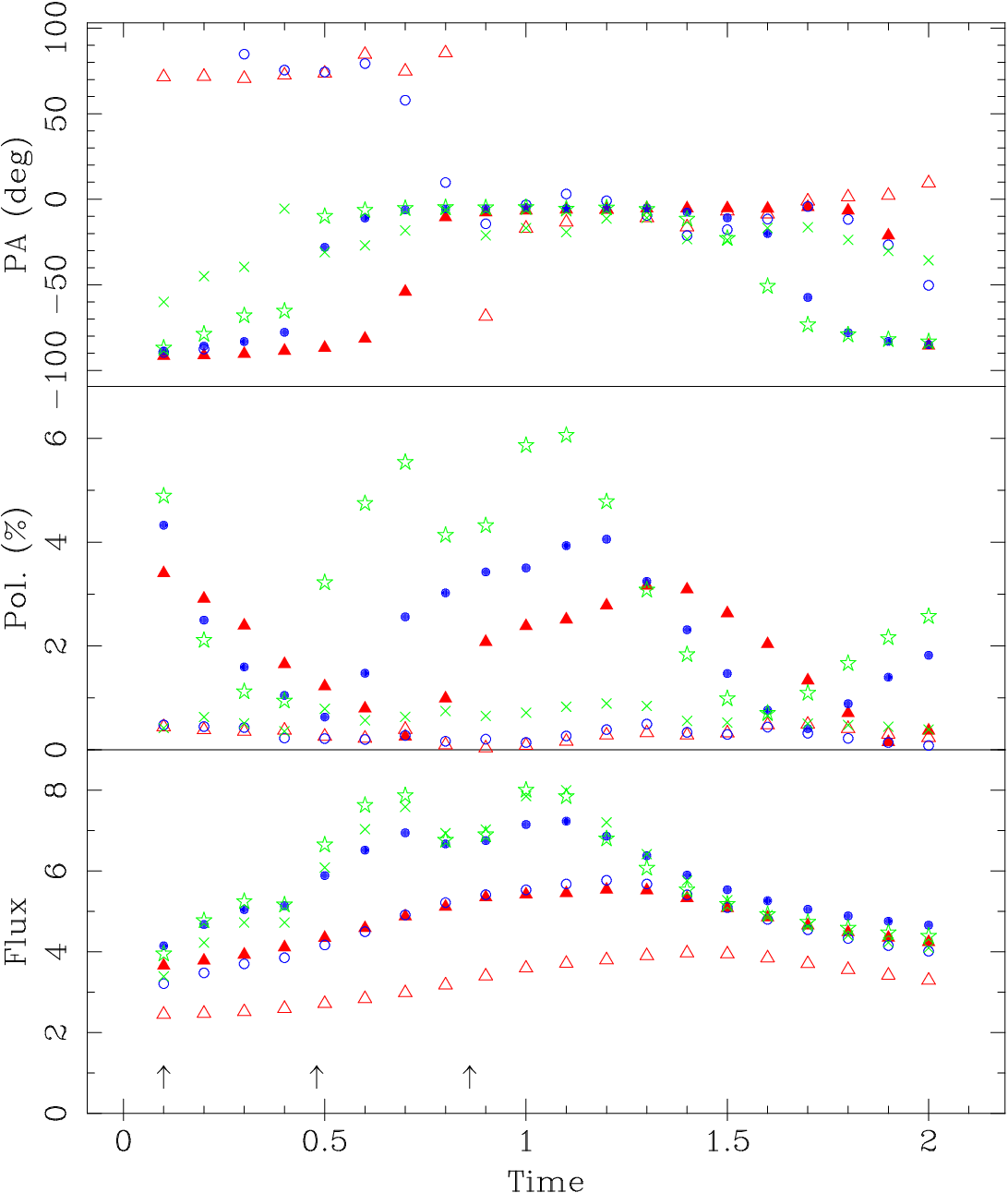}
\caption{The model for 0420-014 at two extremes of the polar viewing angle:
$\theta=1.6$ and $5.6\arcdeg$. The symbols are as for Figure~\ref{fig1}.
An mpeg animation showing in detail the variation of the
light curves with change in model parameter is available online.
}
\label{fig4}
\end{figure}

\subsection{Polar Viewing Angle}\label{polarview}
The parameter sequence shown in Figure~\ref{fig4} (animation online)
spans the range $1.6\le \theta \le 5.6\arcdeg$, and is based on the model
for 0420-014, for which a value of $4.0\arcdeg$ was adopted for the polar
viewing angle. There is a non-negligible change in the character of the total
flux as the viewing angle changes, but this is almost entirely associated
with the fact that the Doppler factor, and thus Doppler shift, change with
viewing angle.  As the angle of view increases, for fixed frequency of
observation, the Doppler factor diminishes leading to a higher emission
frame frequency, a less optically thick flow, and a flatter spectrum. As
noted above, when exploring a range of flow Lorentz factors the opacity
was adjusted to compensate for this effect. This was essential, as without
it the almost two orders of magnitude change in opacity, a completely
optically thick flow would have appeared over most of the parameter range
explored. Here the small range in Doppler factors means that it is not
essential to compensate for its change, allowing a direct exploration of
changes in polar viewing angle.

As noted above, the Lorentz factor of the shocked flow in the 0420-014 model
is $\sim 8$, for which the critical angle is $\sim 7\arcdeg$. The range of
viewing angles thus spans the inner part of the critical cone, the degree
of polarization rising rapidly -- to $\sim 6\%$ at the highest frequency --
as the viewing angle approaches the critical value at which the percentage
polarization is a maximum for a compression transverse to the flow. Note
that the flow does not become appreciably optically-thin at any angle,
so the change in degree of polarization is largely a geometric effect.

The two primary shocks in the 0420-014 model have almost identical
compressions of $\kappa=0.65$, for which the maximum polarization is $\sim
26\%$ \citep{hug85} in the simple case of an optically-thin, compressed
flow (and for the maximum polar angle explored, and given shocked flow
speed, is only marginally less, $\sim 24\%$). In this case, and for
almost all UMRAO sources, that the observed percentage polarizations are
substantially less, rarely above 10\% \citep{all03}, is attributable to a
combination of opacity and cancellation from a modest, but non-negligible
axial field component. Certainly, picking a data point at random from a
source at random will almost invariably yield a single digit percentage
polarization, and it might be thought that models cannot expect to do better
than predict such single digit percent polarization. However, at least
as long as one looks over a restricted enough time interval, for example
a time spanning a few flares, sources have a typical and distinct range
of this characteristic. For example, the periods containing the modeled
flares in 0420-014, OJ~287 and 1156+295 (Paper~I) display values between 1
and 4\%, 0 and 9\%, and 1 and 5\%, respectively. Thus as a rule, and not
just for our modeled sources, there is distinctive behavior that models
can aspire to match quantitatively. As seen in the parameter sequence,
the model percentage polarization is very sensitive to viewing angle,
so this is a particularly well-constrained parameter of the model.

\begin{figure}
\includegraphics[scale=0.45,clip=true]{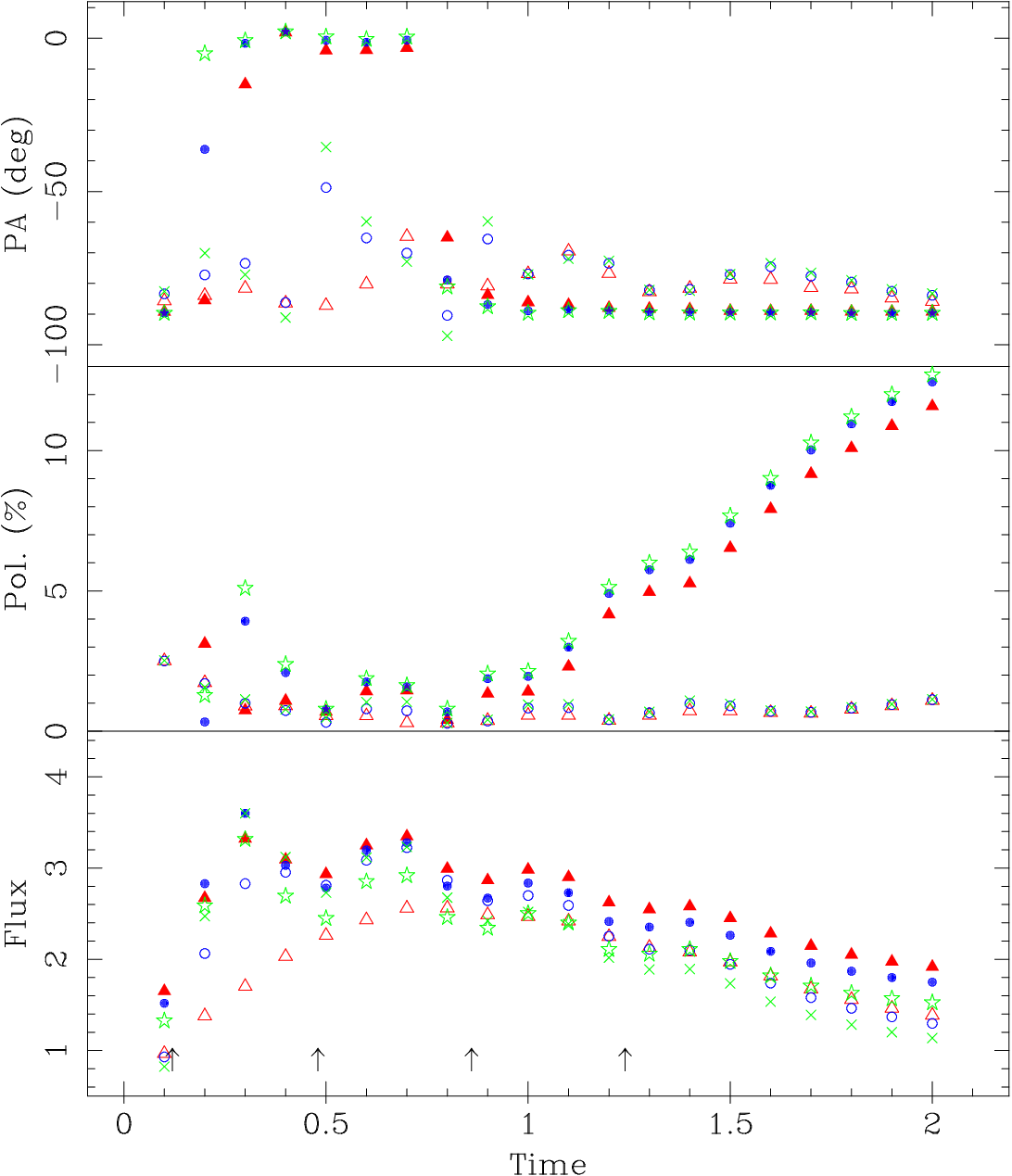}
\caption{The model for 1156+295 at two extremes of the polar viewing angle:
$\theta=0.5$ and $3.5\arcdeg$. The symbols are as for Figure~\ref{fig1}.
An mpeg animation showing in detail the variation of the
light curves with change in model parameter is available online.
}
\label{fig5}
\end{figure}

For comparison, the parameter sequence shown in
Figure~\ref{fig5} (animation online) spans the range $0.5\le \theta
\le 3.5\arcdeg$, and is based on the model for 1156+295 for which a
value of $2.0\arcdeg$ was adopted, and which has a more substantial
axial magnetic field: $\bar{B_z}=0.7$. The quiescent flow has Lorentz
factor $\gamma_f=10.0$, for which the critical angle is $5.7\arcdeg$,
so the parameter sequence probes angles within the critical cone of that
flow. At late times emission from the quiescent flow begins to dominate,
and the percentage polarization rises above $10\%$ for the larger angles
in the sequence because of the strong axial field, seen during quiescence
at $\sim 40\arcdeg$ to the plane of the sky in the emission frame. We
attribute the fact that UMRAO data rarely display such high levels of
polarized emission to the fact that, as noted in \cite{hug11}, truly
quiescent states are themselves a rarity.

\begin{figure}
\includegraphics[scale=0.45,clip=true]{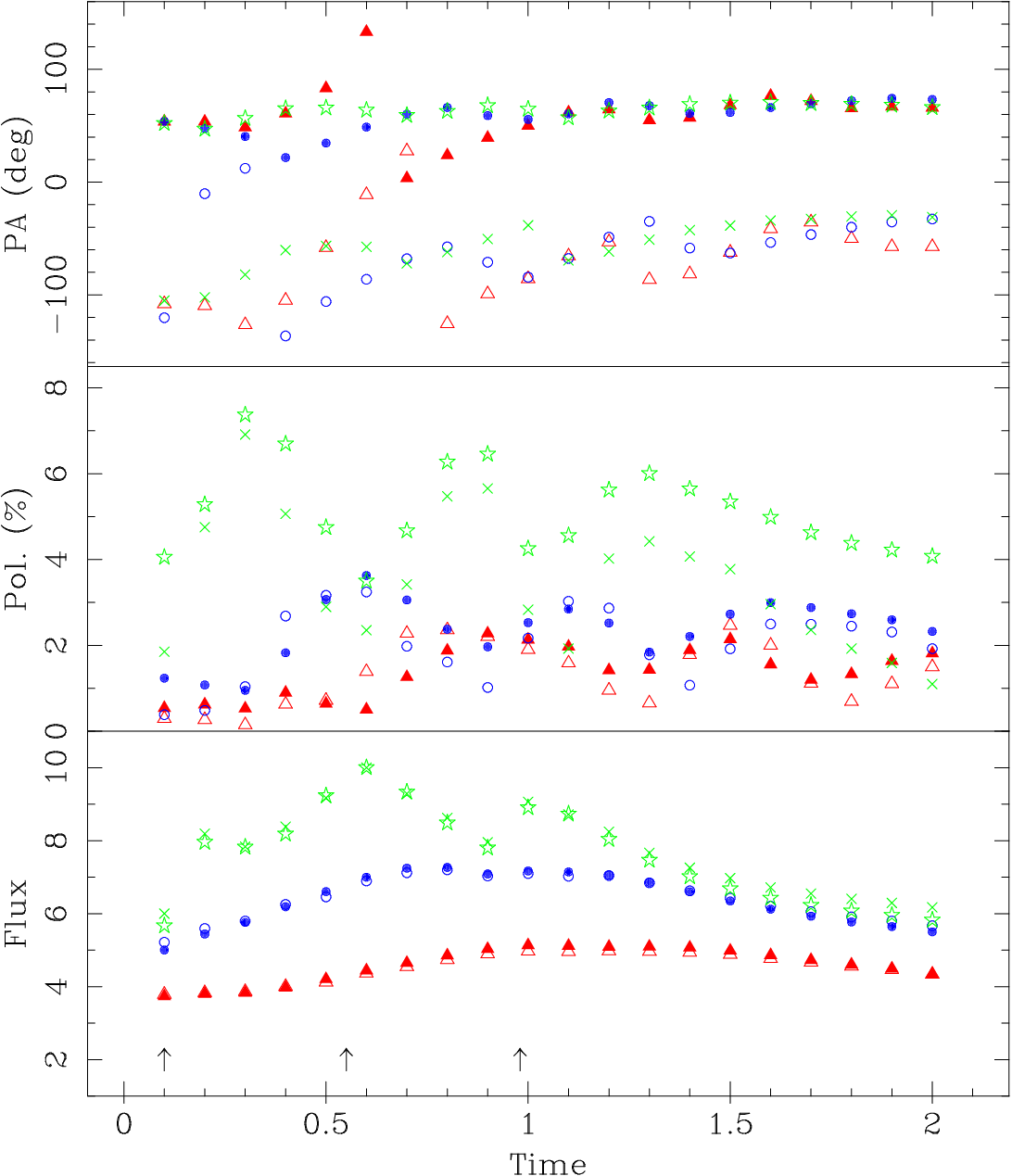}
\caption{The model for OJ~287 at two extremes of the polar viewing angle:
$\theta=0.5$ and $2.5\arcdeg$. The symbols are as for Figure~\ref{fig1}.
An mpeg animation showing in detail the variation of the
light curves with change in model parameter is available online.
}
\label{fig6}
\end{figure}

The parameter sequence shown in Figure~\ref{fig6} (animation online)
spans the range $0.5\le \theta \le 2.5\arcdeg$, and is based on the model
for OJ~287 for which a value of $1.5\arcdeg$ was adopted. Of the three
sources modeled, this has the smallest angle of view, so by bracketing
this value, given that the flow speeds of all the modeled events are
comparable, we are exploring well within the critical cone of the flow,
and thus a small range of Doppler factor.  The range of Doppler factors
never moves the model out of the optically thick domain, and as the peak
flux of all models is scaled to the same value, the total flux light
curve shows little change with $\theta$.  However, while displaying only
a slightly higher opacity than 0420-014, in the case of OJ~287 there is
little change in the percentage polarization; the percentage polarization
of the first component of the modeled flare remains $\sim 7\%$, while
that of the second and third components varies between $4$ and $6\%$. The
shocks in this model are oblique, with $\eta=30\arcdeg$. Depending on the
direction of the shock normal, and azimuth of the observer, a change in
polar angle can have a similar effect to that in the purely transverse case,
can lead to an emission frame angle closer to the plane of compression,
or can lead to an angle further from that plane, in which small angular
changes have little effect (see Section~\ref{aziview}).  For significantly
oblique shocks the percentage polarization light curves do not provide an
unambiguous measure of the viewing angle with respect to the flow axis,
and the viewing angle is less well constrained than for transverse shocks.

\begin{figure}
\includegraphics[scale=0.45,clip=true]{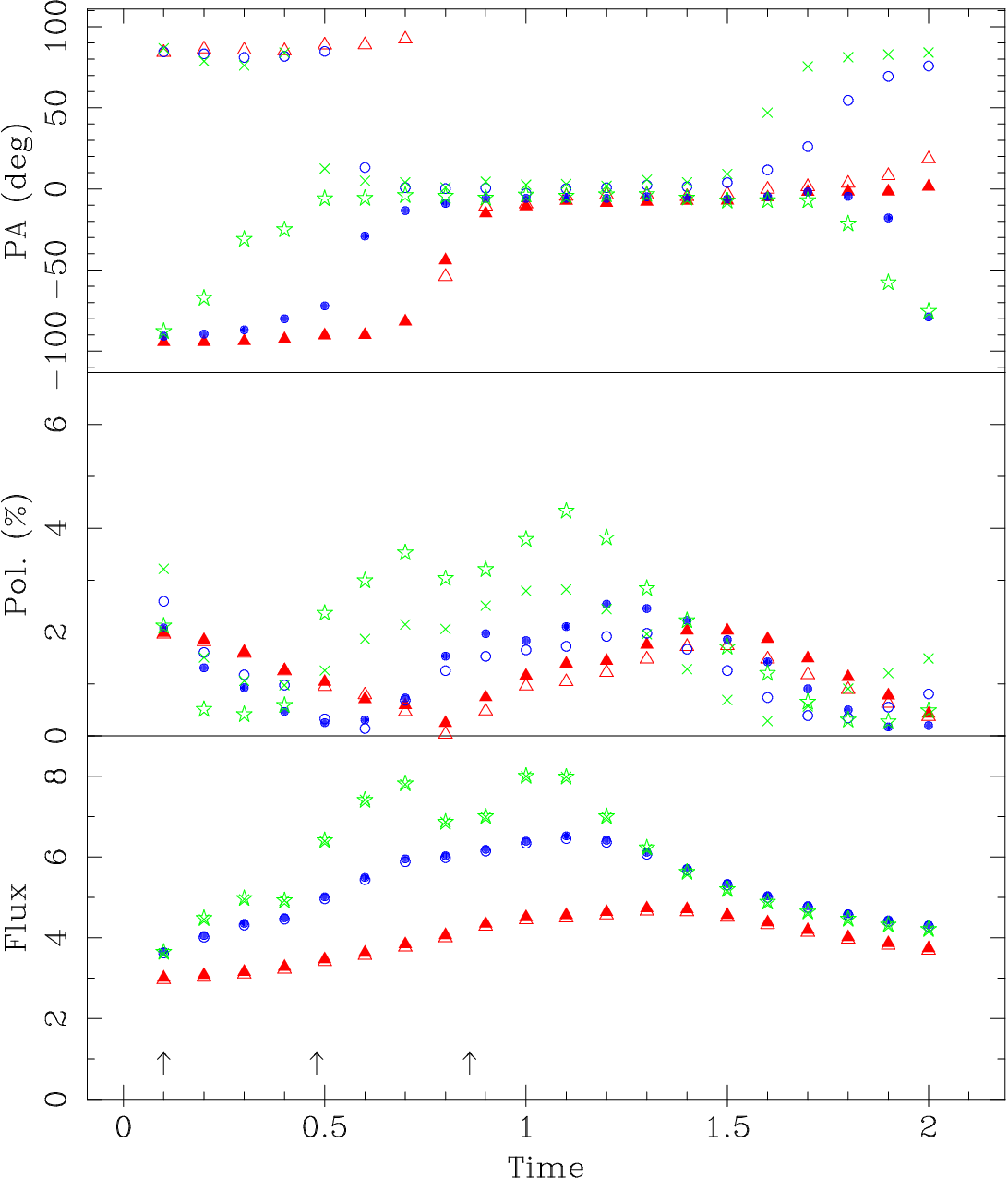}
\caption{The model for 0420-014 at two values of the azimuthal viewing
angle: $\phi=0$ and $120\arcdeg$. The symbols are as for Figure~\ref{fig1}.
An mpeg animation showing in detail the variation of the
light curves with change in model parameter is available online.
}
\label{fig7}
\end{figure}

\subsection{Azimuthal Viewing Angle}\label{aziview}
The parameter sequence shown in Figure~\ref{fig7} (animation online) spans
the range of $\phi$ in $30\arcdeg$ increments, and is based on the model for
0420-014, for which a value of $\phi=90\arcdeg$ was adopted. The figure plots
the model for $\phi=0$ and $120\arcdeg$ to show the full range of behavior
exhibited. The shock has been offset by $1\arcdeg$ from purely transverse,
to illustrate how a change of azimuthal viewing angle, which would play
no role for purely transverse shocks except for subtle effects associated
with particular realizations of the random magnetic field component,
changes the model light curves. This therefore quantifies the role of an
uncertainty in the shock obliquity. Changes are modest but not negligible,
being primarily an increase in the flare percentage polarization by a
few percent across the sequence. Recall that the 0420-014 model adopted
a polar viewing angle of $4\arcdeg$. The offset from a purely transverse
shock structure means that the shocked flow is effectively seen at a polar
angle between $3$ and $5\arcdeg$, depending upon the observer's azimuth,
and indeed the light curves of this sequence are similar to those discussed
in Section~\ref{polarview}, in the range $3.0\le \theta \le 5.0\arcdeg$.
While the modeling well-constrains the polar viewing angle (at least for
transverse shocks), the angle is better thought of as the angle to the
shock normal, rather than to the flow axis.

\begin{figure}
\figurenum{8a}
\includegraphics[scale=0.45,clip=true]{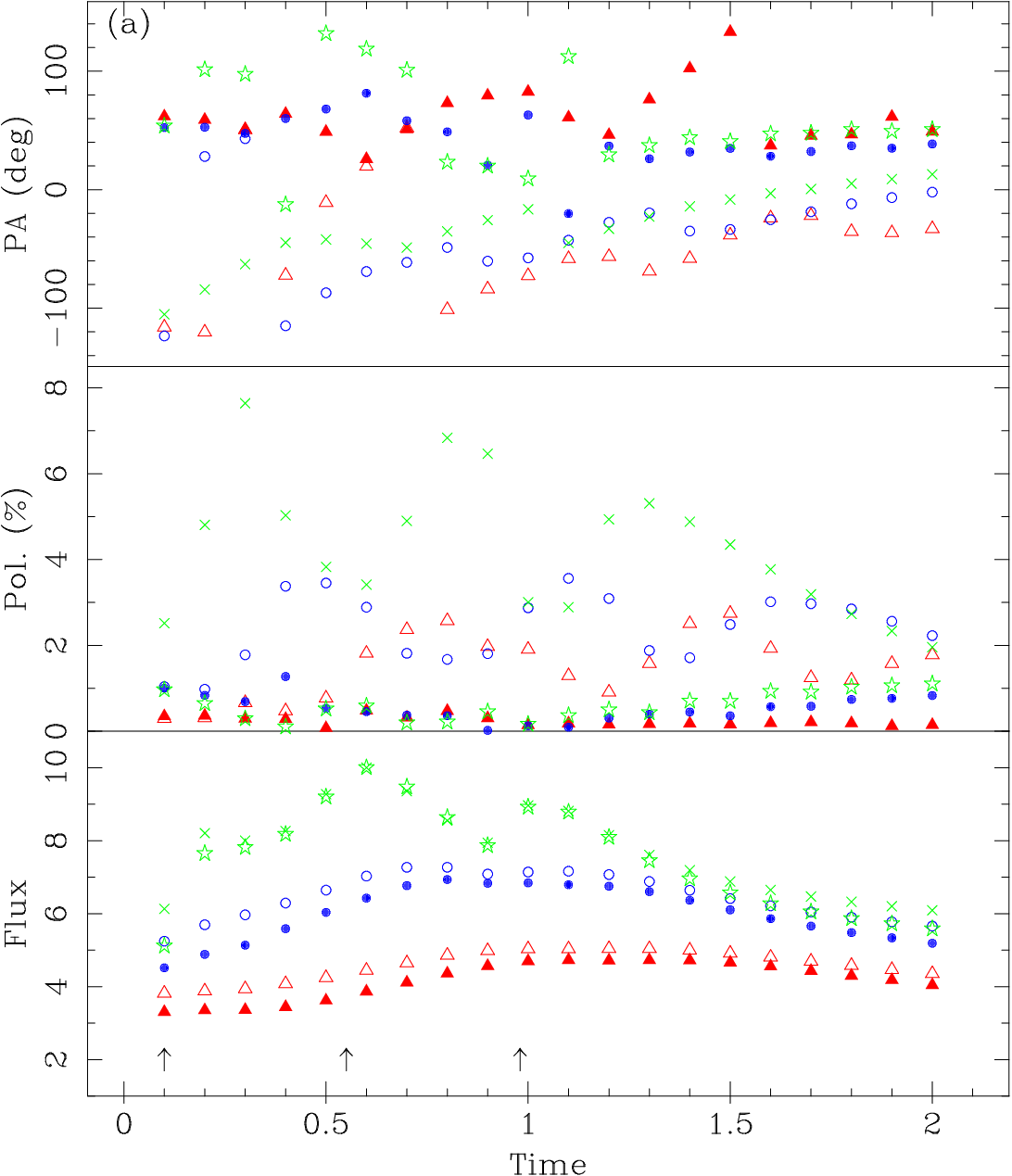}
\end{figure}
\begin{figure}
\figurenum{8b}
\includegraphics[scale=0.45,clip=true]{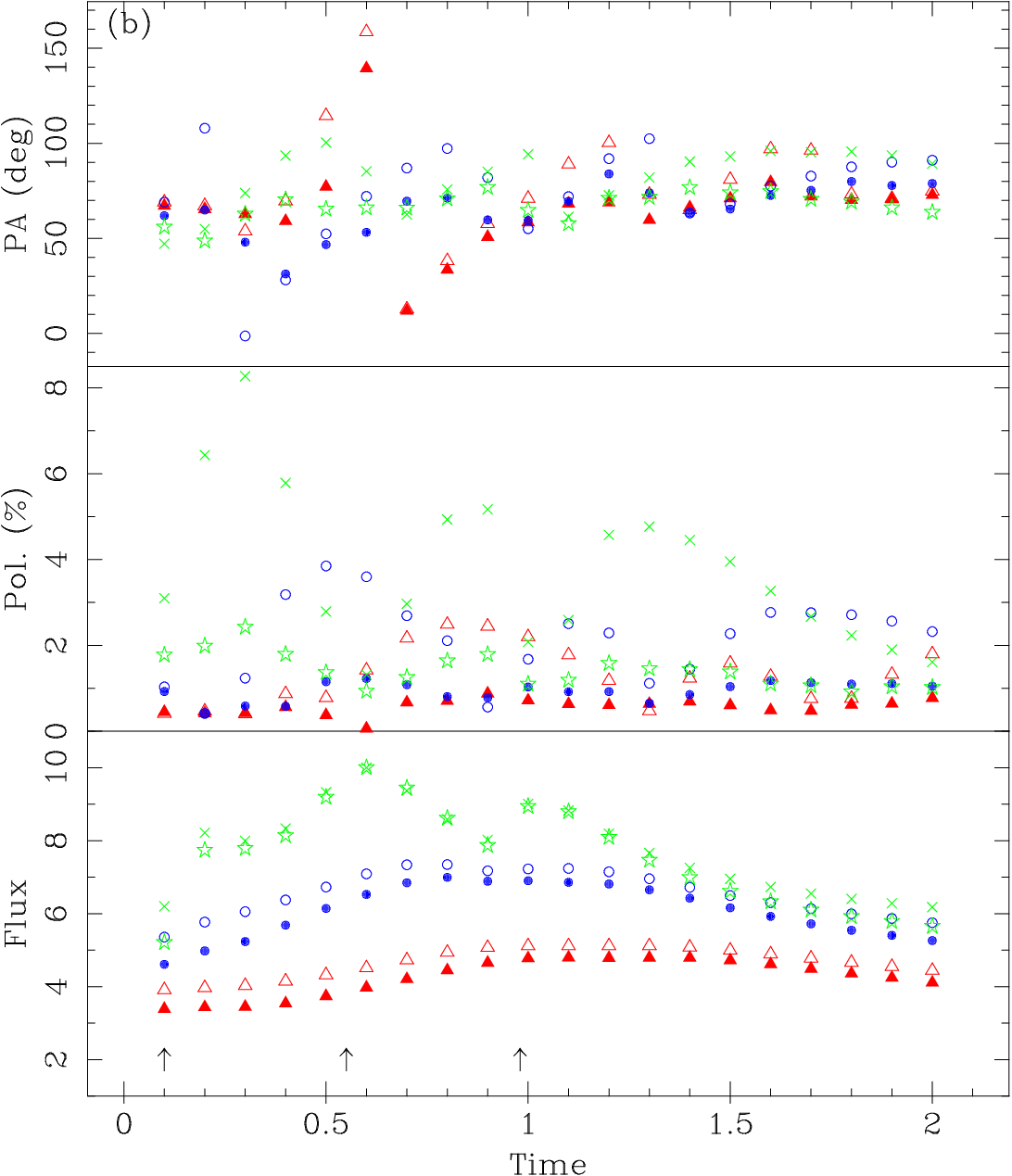}
\end{figure}
\begin{figure}
\figurenum{8c}
\includegraphics[scale=0.45,clip=true]{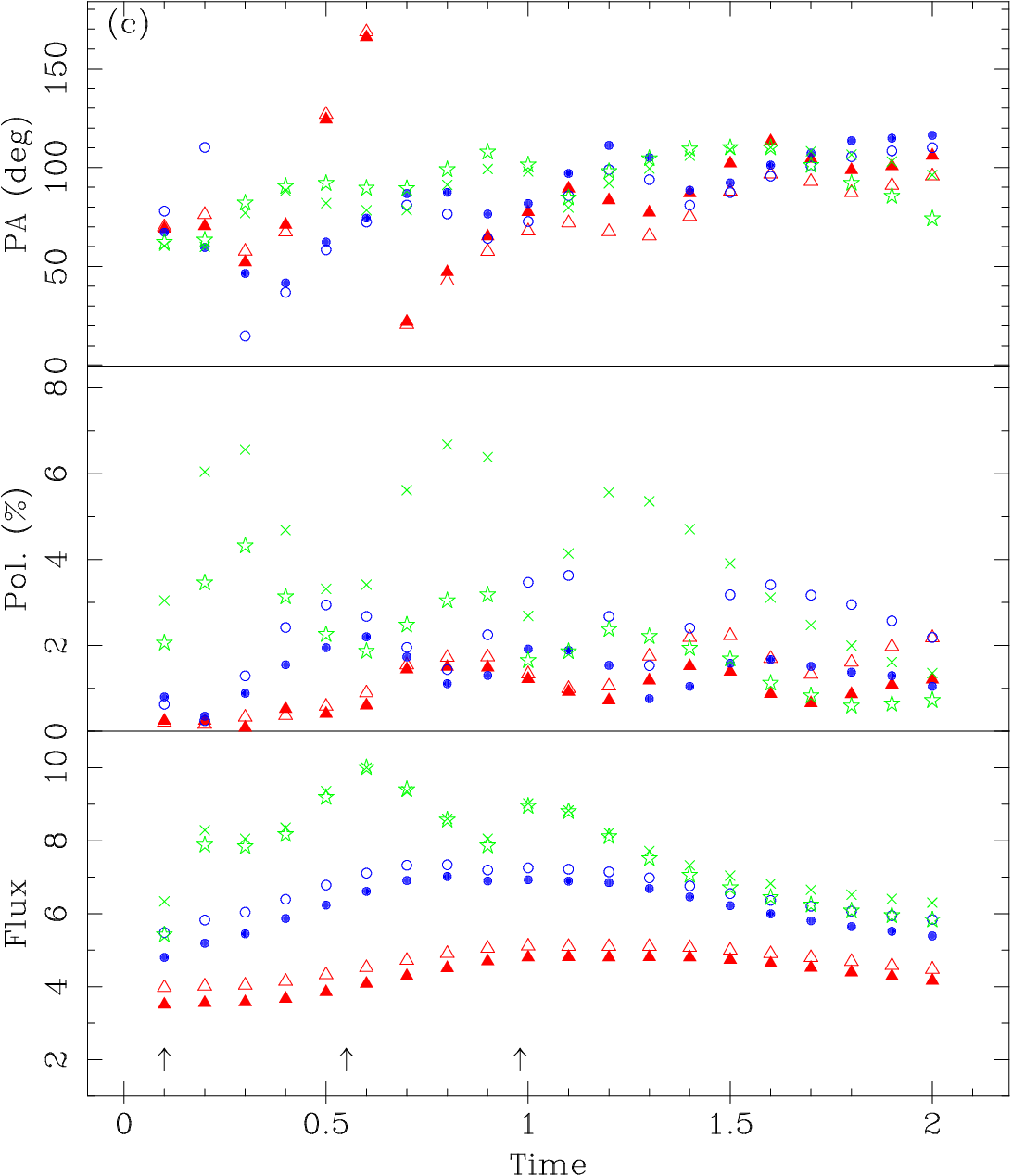}
\end{figure}
\begin{figure}
\figurenum{8d}
\includegraphics[scale=0.45,clip=true]{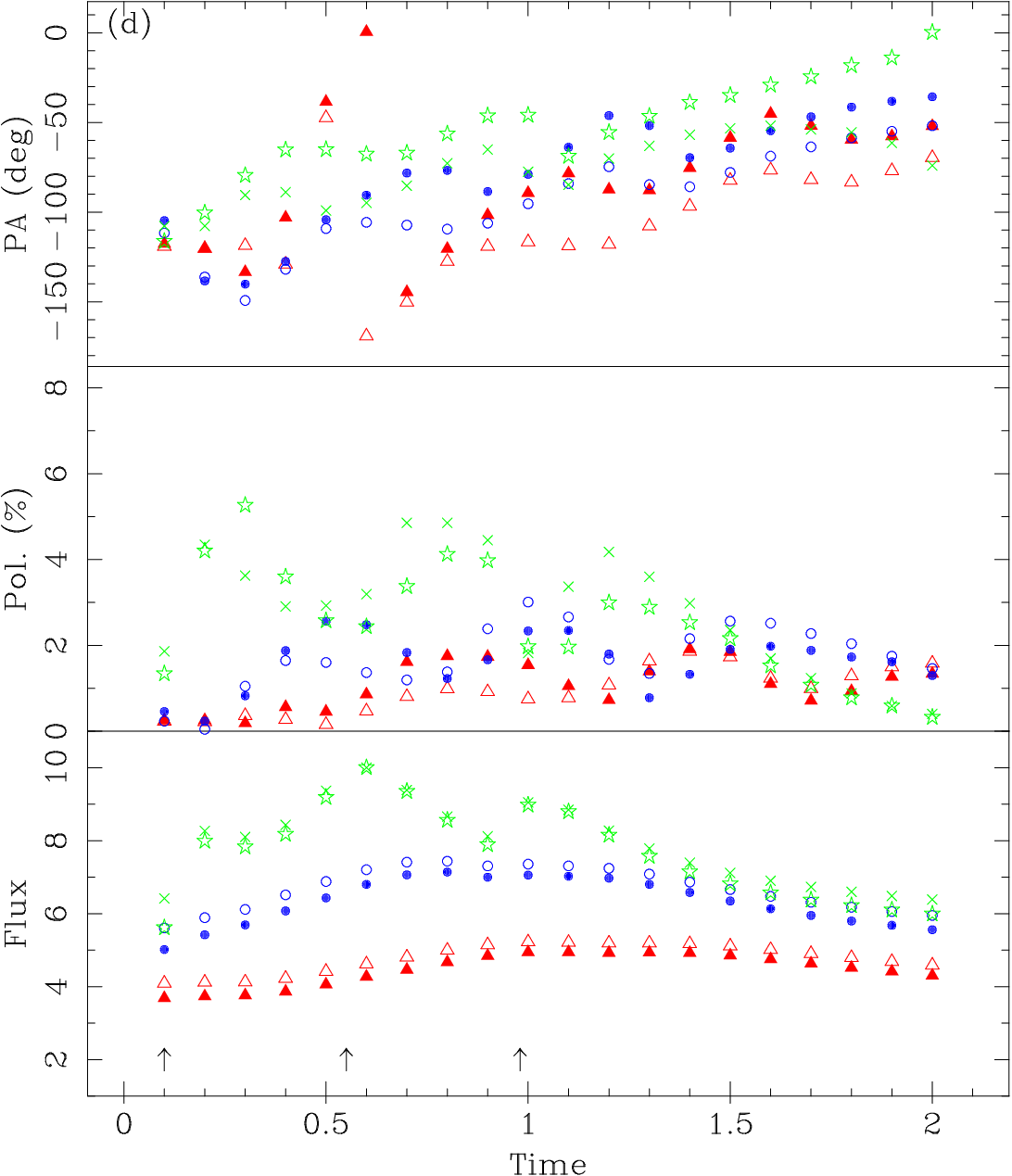}
\end{figure}
\begin{figure}
\figurenum{8}
\includegraphics[scale=0.45,clip=true]{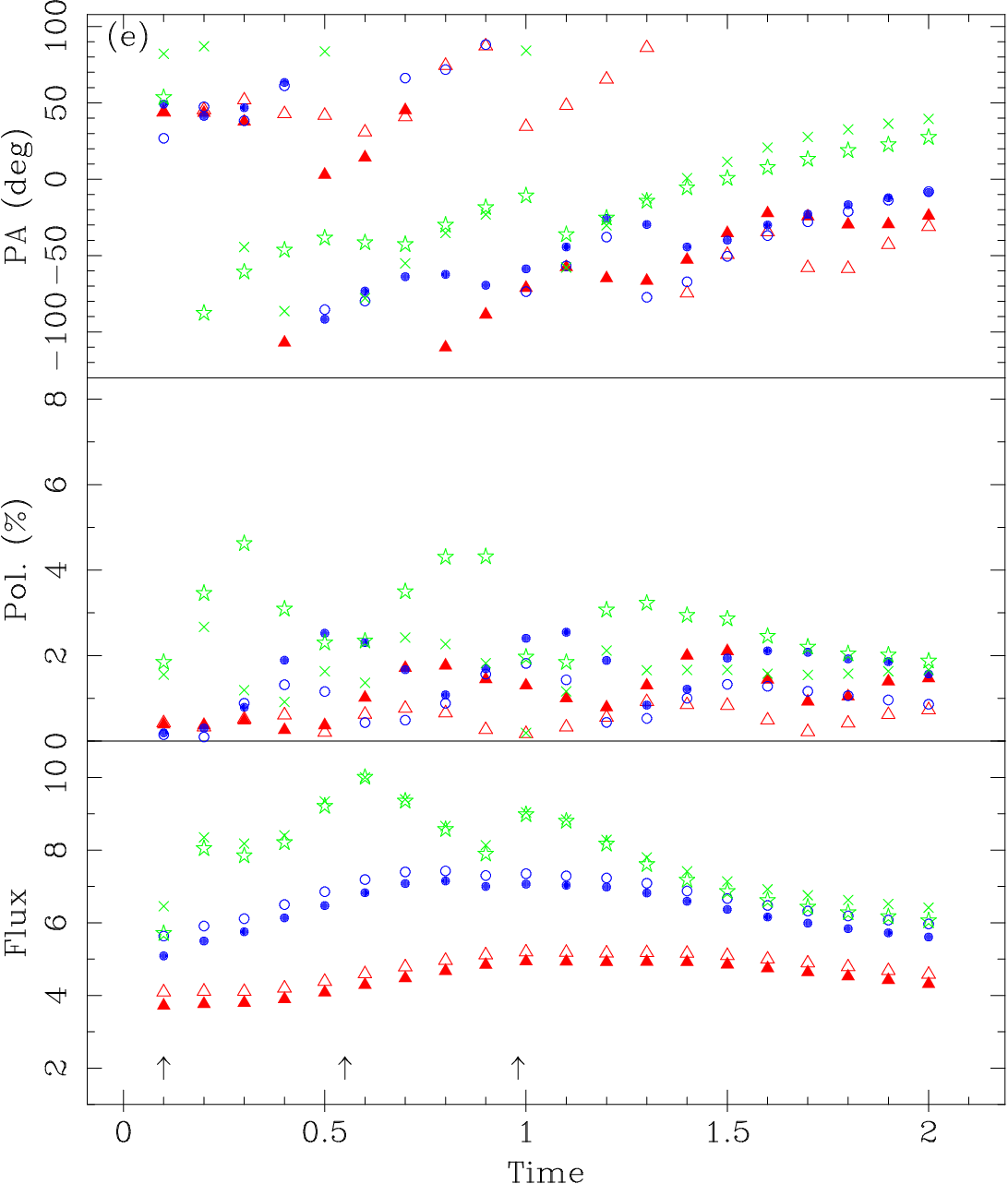}
\caption{The model for OJ~287 at two extremes of the obliquity: $\eta=16$
and $56\arcdeg$. Panels (a)-(e) show the cases of $\phi=0,45,90,135$, and
$180\arcdeg$. The symbols are as for Figure~\ref{fig1}.  
mpeg animations showing in detail the variation of the light curves with
change in model parameter are available online.
}
\label{fig8}
\end{figure}

The parameter sequences shown in Figure~\ref{fig8} (animations online) span a
range of shock obliquities $16\arcdeg \le \eta \le 56\arcdeg$ for each of the
five azimuthal viewing angles $\phi=0,45,90,135$, and $180\arcdeg$, and is
based on the model for OJ~287 for which $\eta=30\arcdeg$, $\phi=45\arcdeg$,
and the shock normal is oriented in azimuthal direction $0\arcdeg$. Given the
direction of the shock normal there is symmetry about the azimuthal direction
$\phi=0\arcdeg$, so we do not display the sequences for $\phi=225\arcdeg$
through $315\arcdeg$. There is a well-defined trend with increasing
azimuthal angle: at $\phi=0\arcdeg$ the peak percentage polarization drops
monotonically from $\sim 8$ to $\sim 0\%$ with increasing shock obliquity,
at $\phi=45\arcdeg$ there is a monotonic drop from $\sim 8$ to $\sim 2\%$,
at $\phi=90\arcdeg$ there is a rise from $6$ to $8\%$, followed by a drop
to $5\%$, and at $\phi=135$ and $180\arcdeg$ the initial, peak and final
polarizations are $5$, $7$, and $6\%$, and $3$, $6$, and $5\%$ respectively.

This variation may be understood as follows.  The compression factor
determines the increase in density and magnetic field across the shock,
and is chosen independently of obliquity. Different obliquities will
yield different upstream and downstream flow speeds in the shock frame,
and thus in the observer's frame; this will lead to different aberrations
for the different cases. The compression is normal to the shock plane,
the orientation of which (with respect to a given observer) changes with
obliquity. For the observer at $\phi=0$ or $180\arcdeg$ the obliquity
determines only the orientation between the (pre-aberration) flow frame
angle of view and the plane of compression: for $\phi=0\arcdeg$, smaller
values of the obliquity angle correspond to a flow frame view close to
the plane of compression, and increasing the obliquity angle moves the
observer farther from that plane, decreasing the percentage polarization;
for $\phi=180\arcdeg$ and small obliquity angle the flow frame view is quite
close to the compression frame, intersects it with increasing obliquity
angle, and with a further increase in the obliquity angle passes onto
the other side of the plane, leading to an initial rise, and then fall in
the percentage polarization. For $\phi=90\arcdeg$ and a transverse shock
($\eta=90\arcdeg$), the angle between the plane of compression and the
(pre-aberration) flow frame angle of view is set solely by the latter,
i.e., polar angle and aberration.  However, for a maximally oblique shock
($\eta\sim 0\arcdeg$) the observer will be in the plane of compression for
all polar angles of view and aberrations (recall that the shock normal points
in the direction $\phi=0\arcdeg$). One might thus expect a monotonic decline
in percentage polarization with increasing obliquity angle, as that takes
the geometry from nearly the maximally oblique case towards the transverse
case. However, the change in obliquity angle also leads to a change in the
downstream flow speed in both the shock and observer's frames, and thus
to a change in aberration. This initially offsets the anticipated trend,
leading to a small initial rise in percentage polarization.

In summary, the peak percentage polarization spans a narrow range of
values as both shock obliquity and observer azimuthal angle are varied
(the constraint on obliquity coming largely from the EVPA behavior),
except for a limited range of azimuthal angles around $\phi=0\arcdeg$
\citep[as noted in][]{hug11}, where the value is close to zero. The azimuthal
angle is thus not well-constrained. The behavior discussed above further
demonstrates that shock obliquity adds uncertainty to the determination
of the observer's polar angle with respect to the flow.

\section{1156+295: The Case Of The Orphan Flare}\label{orphan}
 As noted in Section~\ref{explore}, most interest lies in the complex
interaction of multiple events (structure in both VLBI maps and in
single-dish light curves suggests that we never see isolated events),
so in the above sections we have focused on the consequences of varying
individual parameters in the context of representative models for observed
radio-band outbursts, rather than single shock events. An additional
question that needs to be addressed is the impact of changing the
fundamental sub-structure of these composite events: the number of shocks,
and their length and compression.

This can be very effectively illustrated by considering the model for
1156+295. The model described in Paper~I is based on four shocks of equal
length (10\% of the quiescent flow), equally separated in start time,
and of monotonically decreasing strength -- compressions of $0.5$, $0.6$,
$0.7$, and $0.8$. A number of characteristics of the radio-band outburst
with an associated $\gamma$-ray flare are well-reproduced by the model,
including the total flux density profile and spectrum, the degree and change
of percentage polarization, and the swing in the EVPA. But, intriguingly,
that radio-band outburst is proceeded by one of broadly similar spectral
and temporal characteristics in both total and polarized flux, but with
no cotemporal $\gamma$-ray event.

\begin{figure}
\figurenum{9}
\includegraphics[scale=0.45,clip=true]{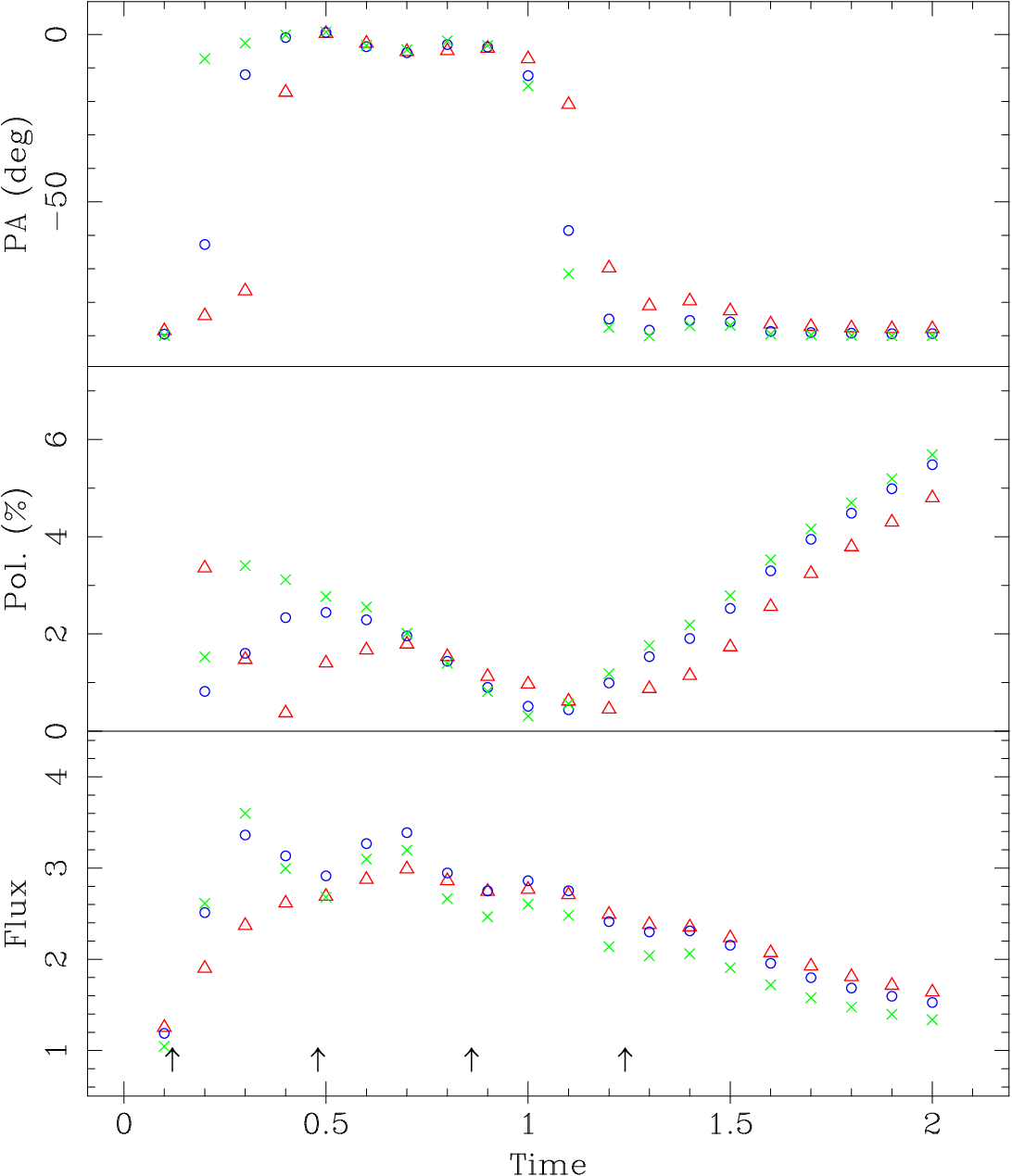}
\includegraphics[scale=0.45,clip=true]{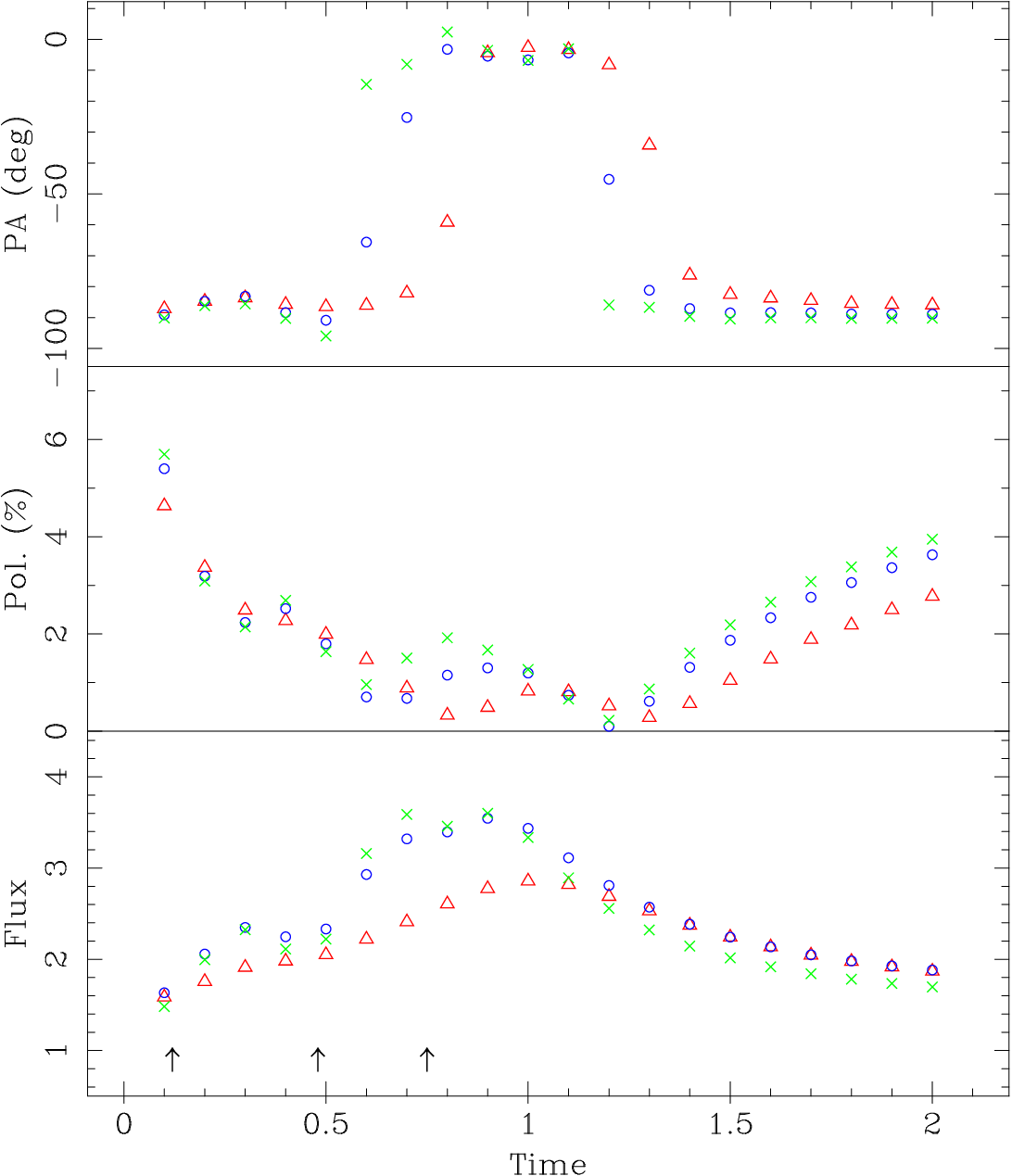}
\caption{Left panel: The model for 1156+295 from Paper~I; this model
corresponds to the radio-band outburst accompanied by an event in the
$\gamma$-ray band. Right: The model for the `orphan flare' exhibited by
the same source -- in this case there is no corresponding $\gamma$-ray
event. Symbols are those adopted in Paper~I. 
}
\label{fig9}
\end{figure}

The `orphan flare' has a total flux light curve similar to those shown
for single shock events in \citet{hug11}, rather than the triangular
form of the second event -- which, together with substructure in the
percentage polarization light curve, was the motivating factor behind
the choice of a four-shock model. That suggests attempting to model the
`orphan flare' with a simpler system of events. Figure~\ref{fig9} shows
both the original model from Paper~I and a model with two shocks, each 10\%
of the quiescent flow in length, and starting at dates 2008.40 and 2008.64.
Their compressions are $0.6$, $0.7$ respectively, corresponding to the
central pair of shocks used to model the second outburst. The same approach
to modeling was used as described in Paper~I, which led to a reduced mean
field strength (30\% as opposed to 50\%) and opacity higher by a factor of
$\times 2.5$, but in all essential respects the only difference between the
models is that of the complexity of the shock system. (A weak `precursor'
event has been included to provide an `initial state' of high percentage
polarization. This is evident in the polarization data just before the
commencement of the flare. There is little evidence from the total flux
light curve of significant additional substructure, and this precursor
could be removed without significantly modifying the model light curves.)

It has been suggested by \citet{bel00} and \cite{ks01} that $\gamma$-ray
bursts (GRBs) can be highly efficient if multiple interactions occur between
the numerous shells that generate internal shocks. The sub-structure of
the $\gamma$-ray flare in 1156+295 that peaks around 2010.5 suggests a
similar scenario. This leads us to hypothesize that the $\gamma$-ray flare
originates from GRB-like interactions upstream of the cm-band outburst region
-- consistent with these events appearing on the rise portion of the radio
band light curve; that these interactions merge some substructures,
and order the flow speeds of those that remain, leading to a sequence of a
few, noninteracting subcomponents to contribute to the radio-band outburst;
and that the absence of a $\gamma$-ray flare during the earlier radio-band
outburst is due to a simpler flow disturbance with much reduced interaction
between components, perhaps because of fewer components as implied by the
radio-band model.

Such a scenario does not address why the converse -- a $\gamma$-ray
flare without an associated radio flare -- is sometimes seen. The radio-band
modeling presented here reveals a significant difference in opacity
between sources, and one possible explanation for the absence of a radio
counterpart to a $\gamma$-ray event is that the event occurs sufficiently
far upstream (nearer the base of the flow) that opacity masks the event
in the cm-band.  By the time the region of accelerated particles
and compressed magnetic field has become partially optically thin in the
radio-band, radiative and adiabatic losses have reduced the emissivity
to a low-enough level that no outburst is evident. A specific variant of
this idea has been suggested by \citet{mac14}. In that model, as a plasmoid
propagates along a jet spine it passes through a ring of shocked material
in the jet sheath.  That ring supplies seed photons for inverse-Compton
scattering by the plasmoid electrons, leading to a rapidly dissipating
$\gamma$-ray flare; model light curves agree well with observations
of a $\gamma$-ray flare seen in the quasar PKS 1510-089. However, much
more theoretical work is needed to understand the complex multi-waveband
behavior of these sources. Also, it is possible that multiple processes
occur, which are not distinguished by current data.

We conclude that modeling as described in this paper has the ability to
discriminate between superficially similar radio-band outbursts, revealing
fundamental differences between them that have the potential to elucidate
important properties of the underlying flows.

\section{Discussion}\label{disc}
 Establishing a well-defined low-energy cutoff to the radiating particle
distribution is significant, because a common feature of many mechanisms
for producing particles of the appropriate energy involves a `heating'
that pushes a subset of low-energy particles into a suprathermal tail
of the energy distribution, followed by an `acceleration' that produces
a power-law distribution from that tail. [See \citet{eh91} for a review
of this, but note that multiple phases of acceleration, for example at
a sequence of oblique or conical shocks, can result in a flat spectrum
with a depleted low energy region, as modeled by \citet{mb13}.] One
would therefore expect the distribution to extend down to the reservoir
of low-energy particles from which the radiating ones derive.  The
discussion of Section~\ref{lowenergycutoff} shows that very low values
of the cutoff are excluded: in only one case (that of OJ~287) is the
cutoff low enough for its effects to be clearly evident in the polarized
flux light curves, but, even there, the cutoff is at highly relativistic
energy; for the other sources, the absence of evidence for such Faraday
effects in the data constrain the cutoff to be even higher. Of course,
that means that in those cases the precise value is ill-constrained, as at
these high values, changes in the cutoff value lead to very small changes
in the light curves, but that does not invalidate the general conclusion
that these flows do not have particle energy spectra extending far below
the energy responsible for the cm-band emission.  The values found in this
analysis are in agreement with the range of values found by \citet{kang14}
(from 5 to 160, with a median of 55). 

The modeling thus provides convincing evidence against the presence of a
low-energy particle distribution, either a reservoir of particles on which
heating and acceleration act, or resulting from entrainment. It appears
that the radiating particles derive not from a low-energy reservoir,
but are produced {\it in situ}, either within the high-energy environment
where the jet is first established, or on larger scales: through tapping
into a Poynting flux or efficient magnetic field line reconnection
\citep{vin14,ss14}. The absence of entrainment is surprising, given the
evidence for a random magnetic field, with the implication that the flow is
turbulent. However, that a significant fraction of the magnetic field energy
is in an ordered component can account for limited transport of particles
transverse to the flow axis. These conclusions illustrate a recurrent theme
of this discussion, namely that many properties of relativistic jets can
be probed effectively only with multifrequency polarization data.

 In Section~\ref{axifield} we further explored the conclusion of Paper~I
that, while a turbulent component of the magnetic field is a crucial
ingredient of the model for UMRAO sources, a significant mean (axial)
field plays an important role in determining the observed polarization
characteristics of the emission. The comparable energy densities of
the random and mean components suggest that the latter arises through a
non-trivial dynamo process, in which growth of the mean field saturates when
the energy densities of the two components are comparable. An alternative
possibility is that local instabilities cause a partial dissipation of the
bulk flow energy, potentially leading to particle heating/acceleration
and turbulent eddies that influence a pre-existing mean field. It might
be thought that a considerable degree of fine-tuning would be needed for
this to occur without total disruption of the flow, but \citet{pk14} have
argued that a rapid expansion causes loss of causal connectivity across
flows of this type (whether relativistic or not, highly-magnetized or
not), allowing the jet spine to exhibit instability, while globally the
jet is stable. Indeed, they note that total disintegration of the jet
spine does not have to be fatal for the integrity of the larger scale
flow. In either scenario -- growth from smaller scales, or disruption
of pre-existing large-scale field -- a significant large-scale field
can persist and inhibit the entrainment of thermal matter across the jet
boundary. As shown by the sequence of simulations in Figure~\ref{fig2},
there is only minor change in the (scaled) total flux light curves as the
relative contribution of the mean field is increased; the polarization data
are crucial for establishing the fraction of magnetic field energy density
in an ordered component, and convincingly fix the value as comparable to,
or a little less than, that of the random component.

In addressing the sensitivity of the modeling to the bulk Lorentz factor
of the flow (Section~\ref{lorentzfactor}) we are confronted with the
complication of the interplay between several of the model parameters.
Broadly, this is a strength of the modeling: for example, the compelling
case for our original model of BL~Lac \citep{hug89b} came from the fact
that the compression able to reproduce the rise in total flux, allowing
for Doppler boosting, was that needed to produce an effective order in
the magnetic field of the shocked flow which, when observed allowing
for the appropriate aberration corresponding to the flow speed and
observer's angle of view to the flow axis, also produced the observed
percentage polarization.  Despite this complication, the sequence shown
in Figure~\ref{fig3} provides a good indication of the sensitivity of
the modeling to the choice of the underlying flow speed. By adjusting the
`opacity parameter' to preserve the spectral character of the total flux
light curve, and maintaining the shock compression across the sequence, the
set of light curves highlight the role played by a change in bulk Lorentz
factor in determining the percentage polarization -- through the change
in Lorentz factor of the shocked plasma for fixed observer-frame spectral
properties and compression, and thus a change in the flow frame viewing
angle for fixed observer angle of view with respect to the flow axis.

It can be argued that an additional parameter change should be `folded in'
to this sequence -- namely, the viewing angle of the observer with respect
to the flow axis -- to establish whether recovering the flow frame viewing
angle of the original model recovers similar light curves in both total
and polarized flux. However, during the iteration process described
in Paper~I, we have seen that as any parameter is adjusted away from its
optimal value, the model light curves deviate from an optimal match to
the data in a systematic way that would not be expected to be offset by
adjustment in one or more of the other parameters. Taken with the fact that
the modeling process leads to a set of observables, such as apparent speed,
consistent with contemporaneous VLBI measurements, although we cannot
formally prove the uniqueness of the model, the modeling process itself
suggests that the model has converged on a unique part of parameter space.
In that context, the goal of the current exploration is limited to formally
demonstrating the model sensitivity to the parameter under discussion --
the bulk Lorentz factor of the flow. Yet again the polarized flux light
curve is crucial in limiting viable values of this parameter: for values
only a little less than that of the model ($\gamma_f=5$), or no more than a
factor of two higher (corresponding to flows with Lorentz factor $<1.5$ on
either side of the model value, as measured in the frame of the model flow)
the predicted percentage polarization is quite different from that observed.
It is encouraging to see how well the modeling is able to narrowly define
a set of parameters describing a particular source at some epoch.

 From the results presented in Section~\ref{polarview} it is clear that,
at least for transverse shocks, the observer's angle of view to the flow
axis is the best-constrained of all the parameters -- a small change in
which leads to very substantial change in the percentage polarization. Being
able to define the viewing angle well, independently of VLBI measurements,
is a strength of this modeling, as knowledge of that parameter is crucial
for the interpretation and modeling of data across the spectrum. In this
context a concern is that the modeling has failed to capture some important
flow physics: we have assumed noninteracting, constant velocity shocks in
rectilinear motion. At least two of these assumptions are readily shown
to be invalid, as extensive VLBI datasets, covering time sequences of a
decade or more, such as MOJAVE \citep{lis13,hom14}, clearly show curvature,
flow acceleration, and, adding yet another layer of complexity, {\it
temporal changes} in properties such as flow curvature. While the spatial
and temporal resolution of VLBI data are not sufficient to convincingly
address the possibility of shock interactions (and if these are occurring
upstream of the 43~GHz core only higher frequency data could probe such
dynamics), our result for the orphan and non-orphan flares in the source
1156+295 (Section~\ref{orphan}) provides evidence that interactions do
occur, and play an important role in establishing the flare properties.

 This does not weaken our conclusions, if it is accepted that the modeling
is establishing parameters at a particular epoch, for a limited section of
a much more extensive flow. While VLBI observations can track components
for much longer times than covered by the single dish light curves/models,
the most pronounced part of a flare in total and polarized flux, as seen
in single-dish light curves, occurs while the disturbance propagates at
most tens of jet radii; over this spatial scale global flow curvature will
play little role, while secular changes in source structure occur over time
scales much longer than that of a single flare. Indeed, the modeling can be
used to explore changes in source orientation and flow speed between flare
epochs. As discussed in Section~\ref{aziview}, we can be less confident that
the modeling has well-defined the angle between the observer's line-of-sight
and the flow axis if the shocks are oblique. Establishing the range of
shock obliquity displayed by sources is itself a useful goal, as it helps
to probe the flow dynamics, and the origin of disturbances. However, from
the perspective of optimally defining flow properties, to probe the relation
between cm-band and $\gamma$-ray flares for example, a case can be made for
selecting activity due to transverse shocks (as suggested by $90\arcdeg$
swings in EVPA); as noted in Paper~I, such cases are quite common, and
the selection of sources for modeling is biased towards these cases.
Additionally, for interpretation and modeling of the $\gamma$-ray data it
needs to be established whether there is significant curvature of the flow
between cm-band and $\gamma$-ray emission regions (the latter upstream of the
former -- see Paper~I), but the modeling provides the observer viewing angle
for the former, and thus the ability to deproject curvature seen upstream
of this region on VLBI maps, if they exist for a given source/flare. Given
the closeness of the observer's line-of-sight to the flow axis in most
cases it seems likely that the actual curvature will prove to be quite small.

 Finally, we reiterate the importance of the polarized flux light curves
for this modeling. It is quite striking how, looking through the first
seven figures of this paper, the opacity, and thus the spectral slope,
vary between models for the three sources addressed here, and the overall
profile and substructure reflect the number, strength and position of the
shocks that contribute to each `flare' -- but ultimately all light curves
are subtle variants on the van der Laan form (see Section~\ref{retarded}).
It is the richness of the polarized flux behavior that provides the
ability to discriminate between parts of parameter space. Existing UMRAO
data, and other datasets, whether existing or yet to be acquired, with
multi-frequency polarimetry that is well-sampled in the time domain provide
a valuable resource for exploring the internal conditions, flow dynamics,
and orientation of the $\gamma$-ray and cm-band emission regions of AGN jets.

\acknowledgements
 This work was made possible by support from NSF grant NSF-0607523, NASA
Fermi grants NNX09AU16G, NNX10AP16G and NNX13AP18G, and by support for
the operation of UMRAO from the University of Michigan.
Andrew Graus provided valuable assistance with the coding.
This research was supported in part through computational resources and
services provided by Advanced Research Computing at the University of
Michigan, Ann Arbor.
We thank an anonymous referee for comments that helped to improve the
manuscript.

\appendix

\section{Retarded Time Effects}\label{retarded}
 The models presented in \citet{hug11}, Paper I, and the current work,
are computed for an instantaneous state of the flow, and do not incorporate
retarded time effects. In this appendix it is shown that while the inclusion
of such effects does influence the multifrequency total and polarized flux
light curves to a discernible degree, the magnitude of the change is not
sufficient to warrant using time-intensive retarded time computations in
the modeling.

 The kinematic models are analytic in the sense that a simple flow geometry
and dynamics are established, assuming a conical jet, and flow-filling
shocks whose downstream parameters are determined from the jump conditions.
However, given the turbulent nature of the flow -- a key ingredient of the
model -- it is necessary to generate realizations of the magnetic field
structure using Monte Carlo techniques, for each time step, for each model.
Thus, although we do not use hydrodynamic simulations as input to the
radiation transfer calculations, the data sets have the complexity of a full
hydrodynamic simulation with an evolved, random magnetic field component.

 The retarded time needed at a given location for radiation transfer through
a particular cell changes with observer orientation, and flow speed (and thus
with flow dynamics, including the number, strength, and obliquity of the
shocks): thus it changes with a change in almost all model parameters. It
follows that inclusion of retarded time effects would impose an enormous
computational burden on an extensive exploration of parameter space,
because a change in any one parameter would require a recalculation of the
retarded state at each location in the flow, at each observer's time step.

\begin{figure}
\figurenum{10}
\hskip 5.0cm \includegraphics[scale=0.45,clip=true]{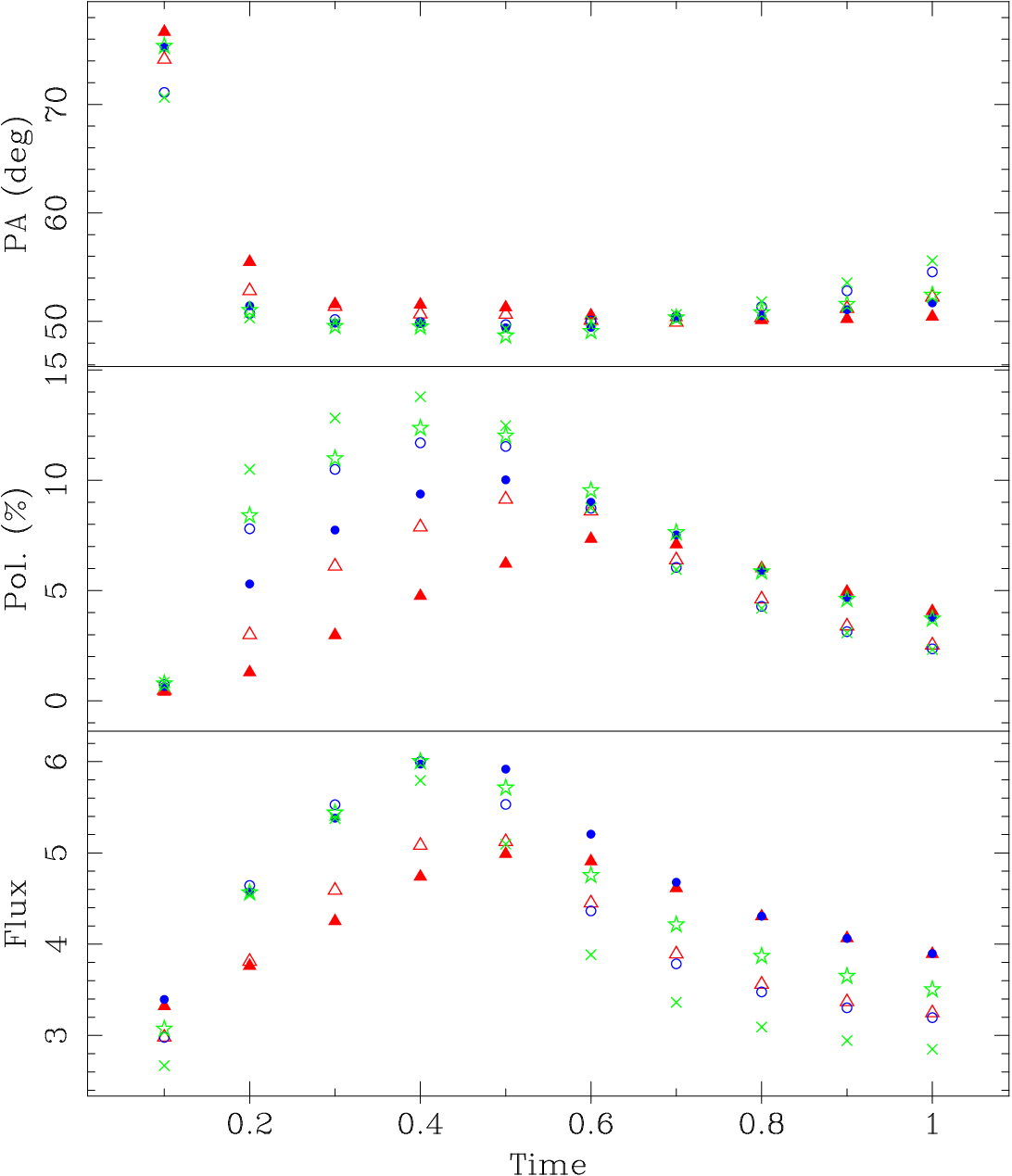}
\caption{Run A from \cite{hug11}, which did not include retarded time
effects, compared with an identical model which does fully account for
such effects, given the $10\arcdeg$ viewing angle of the observer. The
original model is shown using unfilled triangles, circles and crosses for
the frequencies corresponding to the UMRAO frequencies of 4.8, 8.0, and
14.5~GHz (the same convention adopted in Paper~I), while the retarded time
calculation uses filled triangles, circles, and solid crosses.  
}
\label{fig10}
\end{figure}

 Run~A of \citet{hug11} has been recomputed with retarded time effects
included, and the result is shown in Figure~\ref{fig10}.  The
original model is shown using unfilled triangles, circles and crosses
for the frequencies corresponding to the UMRAO frequencies of 4.8, 8.0,
and 14.5~GHz (the convention adopted in Paper~I), while the retarded time
calculation uses filled triangles, circles, and solid crosses. Recall
(Section~\ref{modparam}) that the time coordinate and total flux are subject
to an arbitrary scaling: the original and retarded time runs have been
scaled independently, as would happen in model-fitting, and do not show the
change in burst duration and amplitude that results from the inclusion of
retarded time. The character of the variation in total and polarized flux
and in the position angle of the polarization vector are largely unchanged,
as might be expected given that time delay is ``stretching'' features
significantly only for angles of view close to the jet axis, but that such
structures are just those subject to strong projection effects. For both
transverse and near-axis views, the propagating structure will be fairly
compact, and the total flux variations will have a van der Laan profile
\citep{vdl66}; indeed, the light curves for the total flux in propagating
shock models since the work of \citet{hug89b} all have this character. The
percentage polarization is established by the compression and flow-frame
viewing angle, independently of retarded time, and fold into the total
flux profile to yield a polarized flux light curve similar to that with
no retarded time included.  In effect, a bounded region of propagating
jet plasma always appears `blob'-like. The only significant changes are
a suppression of the flux at the lowest frequency (with a concomitant
reduction in the percentage polarization) due to the longer optical path
length, and increased opacity, and a slower fall in flux late in the event,
again because of the effective extension of the flow along the line-of-sight.

 Quantitatively, the most significant change in total flux occurs near peak
outburst but is a reduction by only $\sim 8$\% at the lowest frequency.
A somewhat larger fractional increase is evident late in the outburst,
but late-time behavior plays little role in model fitting to data. The
concomitant reduction is percentage polarization is only by $\sim 1$\% at
the highest frequency, which is the part of the spectrum most important
for model fitting; while the reduction in percentage polarization is by
$\sim 3$\% near peak outburst at the lowest frequency, the percentage
polarization at 4.8~GHz for a typical source is low, and also plays only a
minor role in the fitting. We conclude that changes in the light curves due
to the inclusion of retarded time are too modest to justify the associated
computational burden, particularly given that other simplifying assumptions
(such as single, noninteracting, constant velocity shocks with rectilinear
motion, but see Section~\ref{disc}) are likely to be at least as important
in limiting our ability to model radio flares.

{}


\begin{thebibliography}{}
\bibitem[Ackermann et al.(2014)]{ack14} Ackermann, M., Ajello, M., Allafort, A., et al.\ 2014, \apj, 786, 157
\bibitem[Agudo et al.(2011)]{agu11} Agudo, I., Jorstad, S.~G., Marscher, A.~P., et al.\ 2011, \apjl, 726, L13
\bibitem[Aller et al.(2003)]{all03} Aller, M.~F., Aller, H.~D., \& Hughes, P.~A.\ 2003, \apj, 586, 33
\bibitem[Aller et al.(2014)]{all14} Aller, M.~F., Hughes, P.~A., Aller, H.~D., Latimer, G.~E., \& Hovatta, T. \ 2014, \apj, 791, 53 (Paper~I)
\bibitem[Beloborodov(2000)]{bel00} Beloborodov, A.~M.\ 2000, \apjl, 539, L25
\bibitem[Blandford \& K\"{o}nigl(1979)]{bk79} Blandford, R.~D., K\"{o}nigl, A.\ 1979, \apj, 232, 34
\bibitem[Cawthorne et al.(2013)]{caw13} Cawthorne, T.~V., Jorstad, S.~G., \& Marscher, A.~P.\ 2013, \apj, 772, 14
\bibitem[Eilek \& Hughes(1991)]{eh91} Eilek, J.~A., \& Hughes, P.~A.\ 1991, Beams and Jets in Astrophysics, 428
\bibitem[Homan et al.(2014)]{hom14} Homan, D.~C., Lister, M.~L., Kovalev, Y.~Y., et al.\ 2014, arXiv:1410.8502
\bibitem[Hovatta et al.(2012)]{hov12} Hovatta, T., Lister, M.~L., Aller, M.~F., et al.\ 2012, \aj, 144, 105
\bibitem[Hughes et al.(1985)]{hug85} Hughes, P.~A., Aller, H.~D., \& Aller, M.~F.\ 1985, \apj, 298, 301
\bibitem[Hughes, Aller \& Aller(1989a)]{hug89a} Hughes, P.~A., Aller, H.~D., \& Aller, M.~F. \ 1989a, \apj, 341, 54
\bibitem[Hughes, Aller \& Aller(1989b)]{hug89b} Hughes, P.~A., Aller, H.~D., \& Aller, M.~F. \ 1989b, \apj, 341, 68
\bibitem[Hughes et al.(1991)]{hug91} Hughes, P.~A., Aller, H.~D., \& Aller, M.~F.\ 1991, \apj, 374, 57
\bibitem[Hughes, Aller \& Aller(2011)]{hug11} Hughes, P.~A., Aller, H.~D., \& Aller, M.~F. \ 2011, \apj, 735, 81
\bibitem[Jones \& O'Dell(1977)]{jod77} Jones, T.~W., \& O'Dell, S.~L.\ 1977, \apj, 214, 522
\bibitem[Jorstad et al.(2001)]{jor01} Jorstad, S. G., Marscher, A. P., Mattox, J. R., et al.\ 2001, \apj, 556, 738
\bibitem[Jorstad et al.(2013)]{jor13} Jorstad, S.~G., Marscher, A.~P., Smith, P.~S., et al.\ 2013, \apj, 773, 147
\bibitem[Kang et al.(2014)]{kang14} Kang, S., Chen, L., \& Wu, Q.\ 2014, arXiv:1409.3233
\bibitem[Kobayashi \& Sari(2001)]{ks01} Kobayashi, S., \& Sari, R.\ 2001, \apj, 551, 934
\bibitem[Kovalev et al.(2009)]{kov09} Kovalev, Y.~Y., Aller, H.~D., Aller, M.~F., et al.\ 2009, \apjl, 696, L17
\bibitem[Lister et al.(2013)]{lis13} Lister, M.~L., Aller, M.~F., Aller, H.~D., et al.\ 2013, \aj, 146, 120
\bibitem[MacDonald et al.(2014)]{mac14} MacDonald, N.~R., Marscher, A.~P., Jorstad, S.~G., \& Joshi, M.\ 2014, American Astronomical Society Meeting Abstracts \#224, 224, \#410.02
\bibitem[Marscher(2006)]{mar06} Marscher, A.~P.\ 2006, Relativistic Jets: The Common Physics of AGN, Microquasars, and Gamma-Ray Bursts, 856, 1
\bibitem[Max-Moerbeck et al.(2014)]{max14} Max-Moerbeck, W., Hovatta, T., Richards, J.~L., et al.\ 2014, \mnras, 445, 428
\bibitem[Meli \& Biermann(2013)]{mb13} Meli, A., \& Biermann, P.~L.\ 2013, \aap, 556, A88
\bibitem[Park et al.(2013)]{park13} Park, K., Blackman, E.~G., \& Subramanian, K.\ 2013, \pre, 87, 053110
\bibitem[Porth \& Komissarov(2014)]{pk14} Porth, O., \& Komissarov, S.~S.\ 2014, arXiv:1408.3318
\bibitem[O'Sullivan et al.(2013)]{osull13} O'Sullivan, S.~P., McClure-Griffiths, N.~M., Feain, I.~J., Gaensler, B.~M., \& Sault, R.~J.\ 2013, \mnras, 435, 311
\bibitem[Rogachevskii \& Kleeorin(2003)]{rk03} Rogachevskii, I., \& Kleeorin, N. 2003, \pre, 68, 036301
\bibitem[Sironi \& Spitkovsky(2014)]{ss14} Sironi, L., \& Spitkovsky, A.\ 2014, \apjl, 783, L21
\bibitem[Valtaoja \& Ter\"asranta(1995)]{vt95} Valtaoja, E., \& Ter\"asranta, H.\ 1995, \aap, 297, L13
\bibitem[Valtaoja \& Ter\"asranta(1996)]{vt96} Valtaoja, E., \& Ter\"asranta, H.\ 1996, \apss, 120, 491
\bibitem[van der Laan(1966)]{vdl66} van der Laan, H.\ 1966, \nat, 211, 1131
\bibitem[Vincent(2014)]{vin14} Vincent, S.\ 2014, International Journal of Modern Physics Conference Series, 28, 60189
\bibitem[Yousef et al.(2008)]{y08} Yousef, T.~A., Heinemann, T., Schekochihin, A.~A., et al.\ 2008, Physical Review Letters, 100, 184501
\bibitem[Zrake(2014)]{zrake14} Zrake, J. \ 2014, arXiv:1407.5626
\end{thebibliography}
\end{document}